\documentclass[floatfix,
 reprint,
 superscriptaddress,
 frontmatterverbose,
 amsmath,amssymb,
 aps,pre
]{revtex4-2}
\usepackage[export]{adjustbox}
\usepackage{xcolor}
\usepackage{graphicx}
\usepackage{dcolumn}
\usepackage{mathtools}
\usepackage{algorithm}
\usepackage{relsize}
\usepackage{mathptmx}
\usepackage{comment}
\usepackage{array}
\usepackage{amssymb}
\usepackage[caption=false]{subfig}
\usepackage[normalem]{ulem}

\graphicspath{{GRAPHS/}}
\newcommand{\eref}[1]{Eq.~(\ref{#1})}

\newcommand{\fref}[1]{Fig.~\ref{#1}}

\begin{document}

\preprint{APS/123-QED}

\title{First-passage and correlated reset-count statistics under threshold and Poissonian resetting in a two-dimensional wedge geometry}

\author{Fazil Najeeb}
\email{fazilnajeeb@cusat.ac.in}
\affiliation{Cochin University of Science and Technology, Kalamassery. P.O,
  Kochi-682022, India}
\author{Arnab Pal}
\email{arnabpal@imsc.res.in}
\affiliation{The Institute of Mathematical Sciences, CIT Campus, Taramani,
  Chennai 600113, India}
\affiliation{Homi Bhabha National Institute, Training School Complex,
  Anushakti Nagar, Mumbai 400094, India}
\author{V.V. Prasad}
\email{prasad.vv@cusat.ac.in}
\affiliation{Cochin University of Science and Technology, Kalamassery. P.O,
  Kochi-682022, India}

\date{\today}

\begin{abstract}
We study the interplay between threshold resetting (TR) and Poissonian resetting in a Brownian target search problem confined to a two-dimensional open wedge of apex angle $\alpha$. The boundary at zero angle acts as an absorbing target, while encounters with the boundary at $\alpha$ trigger TR, instantaneously restarting the search from its initial position. For TR alone, using the renewal framework developed by Biswas \textit{et al} in Phys. Rev. Lett. 135, 227101 (2025), we obtain exact first and second moments of the first-passage time (FPT), whose density exhibits an algebraic long-time tail $t^{-1-\frac{\pi}{2\alpha}}$. To combine the two resetting mechanisms, we thereafter develop a generic random-variable-based renewal formalism to derive the combined mean FPT. We show that the coefficient of variation of the underlying TR dynamics determines whether Poissonian resetting accelerates or hinders the search for $0<\alpha<\pi/4$, whereas for $\alpha\geq\pi/4$, Poissonian resetting always reduces the MFPT. 
Furthermore, we derive the joint count statistics of the two resetting mechanisms and characterize their mutual dependence through Pearson's correlation coefficient. Interestingly, we uncover a crossover from weak to strong correlations as the Poissonian resetting rate is varied, with the crossover behavior exhibiting a nontrivial dependence on the initial search position. Our framework provides a basis for studying competing resetting mechanisms in more general geometries.
\end{abstract}

\maketitle

\section{Introduction}
First-passage time (FPT) problems are  quite important in the context of random
processes, providing natural timescale in which a stochastic particle first encounters a target or threshold~\cite{redner2001guide,schuss2013brownian}. Applications
range from transcription-factor binding, reaction-diffusion
kinetics, intracellular
signaling, to transport in disordered environment~\cite{redner2001guide,metzler2014first,condamin2007first}.
Geometry of the domain serves an important role in determining the mean first passage times in search processes. For example,  a random walker may perform large excursions in finding out a target in a non-compact domain resulting in a mean first passage time that is infinite. In fact, the first passage distribution is well characterized for a wide class of processes in non-compact domains.~\cite{redner2001guide,bray2013persistence,metzler2014first}.  A physically relevant situation, probably, is one of confinement:
many search processes unfold inside cells, pores, and cavities,  where
reflecting walls, absorbing boundaries, and geometric constraints
reshape the FPT distribution in ways  
the former does not
capture~\cite{schuss2013brownian,holcman2014narrow}.
 
Within this context, the wedge -- a two-dimensional domain bounded
by two absorbing walls meeting at an interior angle $\alpha$ -- occupies a position of particular mathematical and physical significance.
For the system which is confined in the angular dimension but open along the radial coordinate the first passage statistics to the wedge boundaries has been carried out in detail,  illustrating first passage statistics to the absorbing wedges 
depending on the wedge parameters~\cite{redner1999capture}.
For a diffusive particle in  two dimensions confined within absorbing wedge boundaries of apex angle $\alpha$, the survival probability has been shown to follow a power law as $t^{-\pi/(2\alpha)}$ with the mean first passage time diverging at a wedge angle $\pi/2$. Analytical form for the first passage distribution for wedge angles $\alpha=\pi/n$ with $n=2$ and for odd values~\cite{dy2008} and for any positive integer values have been derived \cite{dy2} which was further extended to any arbitrary values of $\alpha$ in~\cite{majumder2015}.

Removing trajectories with large excursions help in decreasing the mean first passage time and a mechanism that is usefully employed for this is the stochastic resetting. Here, the
particle is returned to its initial position at a constant Poissonian
rate $\lambda$~\cite{evans2011diffusion,evans2020stochastic} which truncates the otherwise heavy tail of the 
first-passage distribution, converting a potentially divergent MFPT
into a finite quantity, which can be tunable to have an optimal rate that
minimizes the mean search
time~\cite{evans2011diffusion,evans2011prl,evans2020stochastic,reuveni2016optimal,pal2017first,kusmierz2014first,pal2016diffusion,pal_first_2019}. For comprehensive reviews of the field, including its recent developments and emerging directions, we refer the reader to Refs.~\cite{evans2020stochastic,pal2022inspection,gupta2022stochastic,pal2024random,nagar2023stochastic}. 

Stochastic resetting is commonly implemented through an externally prescribed random mechanism, whereby the system is returned to a reference configuration at times that are statistically independent of its dynamical state. A fundamentally different situation arises when the resetting mechanism is coupled directly to the evolution of the system. In such cases, the occurrence of a reset is determined by the system reaching a specified threshold in an internal variable, rather than by an independent timing process. This class of state-triggered protocols is known as \textit{threshold resetting} (TR)~\cite{Biswas2025_Threshold,biswas2026optimal,das2026non,das2026freezing}, or equivalently, \textit{first-passage resetting}~\cite{debruyne2020optimization,biroli2026first}. An essential feature of TR is that the resetting times inherit the temporal and statistical correlations of the underlying dynamics, since they are determined by the very trajectories being interrupted. Such threshold-mediated resets arise naturally in various 
systems; the membrane potential of an integrate-and-fire neuron is
reset to its resting value each time it crosses the firing
threshold~\cite{burkitt2006review}.

\begin{figure*}[htbp]
  \centering
  \subfloat[Pure threshold resetting.\label{fig:threshold_schematic}]{%
    \includegraphics[width=0.485\textwidth]
      {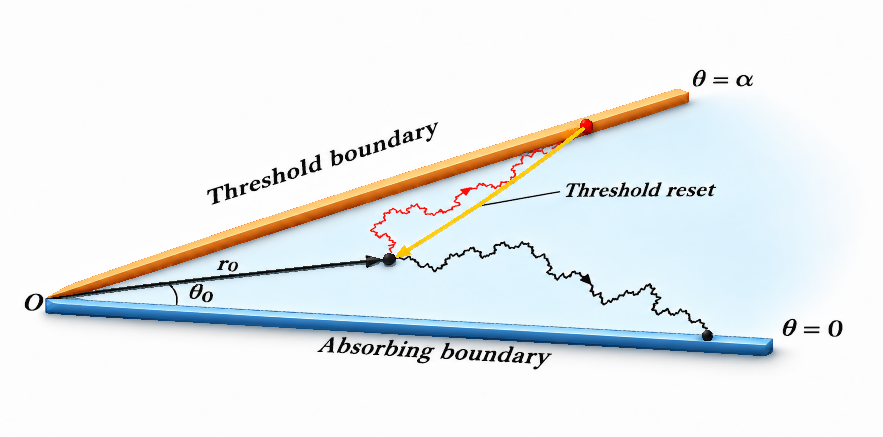}%
  }%
  \hfill
  \subfloat[Combined threshold and Poissonian resetting.\label{fig:combined_schematic}]{%
    \includegraphics[width=0.485\textwidth]
      {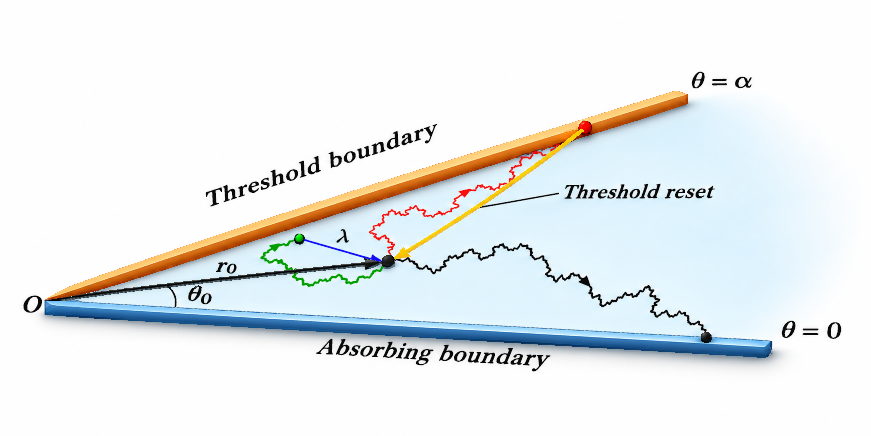}%
  }%
  \caption{Schematic of target search in a 2D wedge under resetting mechanism. In both panels, the ray
  at $\theta=0$ is assumed to be an absorbing boundary, while the ray at $\theta=\alpha$
  is referred to as a threshold boundary. Panel (a): When the particle reaches the threshold,
  it is instantaneously returned to $(r_0,\theta_0)$, as shown by the
  yellow arrow. This is called threshold resetting. Panel (b): In addition to (a), the particle can also be 
  reset to the same point, independently,
  at a Poissonian rate $\lambda$, as shown by
  the blue arrow.
  }
  \label{fig:reset_schematics}
\end{figure*}

Research on threshold resetting (TR) has gained momentum only in recent years, with investigations spanning a variety of single- and many-particle settings. The effect of threshold resetting has been studied in detail in one-dimensional geometries~\cite{debruyne2020optimization,Biswas2025_Threshold,biroli2026first,biswas2026optimal}. One of the early studies by De Bruyne \textit{et al.} examined a diffusing particle confined to a finite domain, focusing on its stationary properties and on optimizing a reset--reward--cost objective function~\cite{debruyne2020optimization}. For a single Brownian particle on a semi-infinite line, resetting upon arriving at a fixed threshold location produces a non-stationary spatial probability distribution, with the mean number of resets grows in time as $\sqrt{t}$~\cite{debruyne2020optimization} in the long-time limit. Alongside these developments, Das \textit{et al} established a general approach to studying non-equilibrium behavior under TR ~\cite{das2026non,das2026freezing}. In particular, it was shown that a finite mean first-passage time to the resetting threshold ensures the existence of a NESS, whose spatial distribution can be expressed in terms of the ratio of the mean local time to the mean first-passage time.  The formalism was then illustrated for drift-diffusion and ballistic random walks showing many interesting relaxation behavior~\cite{das2026non,das2026freezing}. 

The extension to many searchers was considered by Biroli \textit{et al.}, who characterized the non-equilibrium steady state, derived the full counting statistics, and uncovered dynamical correlations in a system of $N$ diffusing particles~\cite{biroli2026first}. 
Another line of investigation has focused on the implications of TR for search efficiency. Considering $N$ searchers, optimization of first passage times to a target was considered in one dimension for a generic searcher dynamics~\cite{Biswas2025_Threshold}. In this context, Biswas \textit{et al.} analyzed ensembles of diffusive and ballistic searchers and identified nontrivial features in the optimization of their first-passage times~\cite{Biswas2025_Threshold,biswas2026optimal}. Subsequent developments have broadened the scope of these studies. Garcia \textit{et al.} considered a single Brownian searcher in one dimension searching for spatially distributed targets under the combined action of threshold and Poissonian resetting~\cite{garcia2026target}.
The role of boundary-triggered resetting in bounded two-dimensional geometries for a single searcher has also been investigated, revealing that the resulting search advantage is controlled by the reset location, the arrangement of interior targets, and the waiting time at the boundary~\cite{chen20262d}.

In a recent work~\cite{3bc4-nlkt}, we investigated Brownian motion in a two-dimensional wedge under Poissonian resetting, with both wedge boundaries acting as absorbing edges. Owing to the open geometry and the eventual absorption of the particle at either boundary, the system does not admit a time-independent stationary state. Instead, we focused on how stochastic resetting modifies the first-passage properties and biases the exit statistics between the two boundaries. In this work, we first investigate threshold resetting in a two-dimensional wedge geometry, which offers a particularly rich setting to explore the interplay between confinement and resetting. Despite having two boundaries, the wedge provides only partial confinement owing to its unbounded radial extent. Consequently, the first-passage time (FPT) distribution exhibits a heavy algebraic tail, with an exponent determined by the wedge angle. Interestingly, by varying this single geometric parameter, the mean FPT can be either finite or divergent, allowing us to explore distinct first-passage regimes within the same geometry. Since threshold resetting is intrinsically governed by boundary encounters, the wedge geometry provides an ideal setting to investigate how geometric constraints shape the resulting search dynamics. Furthermore, by incorporating spatially independent Poissonian resetting, we examine the combined effect, focusing on its consequences for search efficiency, MFPT optimization, and the correlations between their respective resetting events. We summarize the exact set-up and key results in the following.

We consider a single Brownian searcher, having a random coordinate $(r(t),\theta(t))$ at a given time $t$, starting at an initial position $\vec r_0=(r_0,\theta_0)$ inside the wedge [see Fig. \ref{fig:reset_schematics}]. The boundary at $\theta=0$ acts as an absorbing target, whereas, the boundary at $\theta=\alpha$,  triggers an instantaneous reset to the initial position $\vec r_0$, from where the search restarts. We first obtain exact expressions for the first and second moments of the first-passage time under TR, highlighting their singular nature for apex angles $\alpha>\pi/2$ and $\alpha>\pi/4$ respectively. Probing the long time behavior of the first-passage time distribution to  the absorbing target, we find that the algebraic tail namely $f_{TR}(t)\sim t^{-(\frac{\pi}{2\alpha}+1)}$ of the underlying non-reset Brownian motion in wedge \cite{redner2001guide,majumder2015}, is still preserved despite the threshold resetting mechanism. 

We then introduce, in addition to TR, an independent Poissonian resetting (PR) operating at a  rate $\lambda$. PR interrupts randomly, the excursions and instantaneously restarts the search from the initial position. To deal the combined effect, we develop a random-variable-based renewal framework and derive the total MFPT. The coefficient of variation ($CV_{TR}$) of the FPT under TR provides a criterion for determining whether Poissonian resetting enhances or hinders the search, with distinct optimization regimes emerging for different wedge angles. For $0<\alpha<\pi/4$, the curve $\mathrm{CV}_{TR}=1$ separates
regions in the $(u,\alpha)$ phase diagram where a small positive reset
rate lowers the MFPT when $\mathrm{CV}_{TR}>1$, or raises it for
$\mathrm{CV}_{TR}<1$. Finally, we investigate the joint statistics of the TR and Poissonian resetting counts and characterize their mutual dependence through Pearson's correlation coefficient. Interestingly, increasing the Poissonian resetting rate does not necessarily strengthen their correlation; instead, the correlation can span the entire range from zero to unity upon varying the resetting position, revealing a nontrivial crossover between weakly and strongly correlated resetting events.

The remainder of the paper is organized as follows. Section~\ref{models} introduces the wedge geometry and the two resetting protocols, and defines the survival probability and boundary currents of the process without resetting. Section~\ref{target} presents the first two moments and the long-time behavior of the first-passage-time density under pure threshold resetting. In Sec.~\ref{combined_fpt}, we derive the mean first passage time combined resetting protocol and use the coefficient of variation to identify when a small Poissonian reset rate reduces the mean search time. Section~\ref{sec:reset_counts} examines the threshold and Poissonian reset counts before target absorption. We calculate their statistical properties such as mean, joint distribution, correlation. We also look at the asymptotic expression for the correlation for  large resetting rate $\lambda$. Section~\ref{sec:conclusion} summarizes the main findings and discusses possible future directions. Supporting derivations are provided in the Appendices.

\section{\label{models}Model and key observables}

Consider a Brownian particle with diffusion coefficient $D$ in the
two-dimensional wedge domain defined in radial polar coordinates as,
\begin{align}
\Omega_\alpha
=
\left\{(r,\theta):r>0,\ 0<\theta<\alpha\right\}.
\end{align}
The particle starts from
$\vec r_0=(r_0,\theta_0)$, where $r_0>0$ and
$0<\theta_0<\alpha$. We use $u=\theta_0/\alpha$, with $0<u<1$, for the fractional angular position of the starting point.

The Brownian trajectory in Cartesian coordinates obeys the
Langevin equation
\begin{align}\label{trajectory_langevin}
d\vec r(t)
=
\sqrt{2D}\,d\vec W(t),
\qquad
\vec r(0)=r_0\bigl(\cos\theta_0,\sin\theta_0\bigr),
\end{align}
where $\vec W(t)$ is a two-dimensional Wiener process. The radial and
angular coordinates used below are defined by
\begin{align}\label{trajectory_polar_coordinates}
\nonumber \vec r(t)
&=
\bigl[r(t)\cos\theta(t),r(t)\sin\theta(t)\bigr],
\\
r(t)&=|\vec r(t)|,
\\ \nonumber
\theta(t)&=\arg\vec r(t).
\end{align}
For the renewal calculations, we use an non-reset process without resetting, in which both angular boundaries are absorbing. In the present TR wedge model, reaching $\theta=0$ terminates the search, whereas reaching $\theta=\alpha$ restarts the search from $\vec r_0$. The statistical properties of target search in the non-reset process, allows to determine the time of the first boundary encounter and survival probability till any arbitrary time. Boundary-resolved quantities help us distinguish the events of target absorption from threshold resetting; in a combined resetting model, an independent reset even may occur before these boundary encounters.
Therefore we start by reviewing a few these statistical quantities without reset. Let us define the first exit time of this non-reset process,
\begin{align}\label{combined_spatial_exit_timeNEW}
    T_0=\min\{t>0:\theta(t)\in\{0,\alpha\}\}
\end{align}
The value of $\theta(T_0)$ identifies the boundary reached. The propagator $P_0(r,\theta,t\mid r_0,\theta_0)$ is the probability density of this non-reset process. We suppress its fixed initial-position arguments. It evolves in the bulk according
to
\begin{equation}\label{diff_eqn}
\frac{\partial P_0}{\partial t}=D\nabla^2P_0,
\end{equation}
where the Laplacian in radial polar coordinates is
\begin{align}\label{polar_laplacian}
\nabla^2
=
\frac{1}{r}\frac{\partial}{\partial r}
\left(r\frac{\partial}{\partial r}\right)
+
\frac{1}{r^2}\frac{\partial^2}{\partial\theta^2},
\end{align}
with the initial condition $P_0(r,\theta,0) = \frac{\delta(r-r_0)\delta(\theta-\theta_0)}{r}$ also subject to the absorbing boundary conditions $P_0(r,0,t)=P_0(r,\alpha,t)=0$. 
Integrating the above propagator over the wedge domain gives the survival probability of the non-reset process
\begin{align}\label{combined_Q0_definition}
Q_0(t)
&=\Pr(T_0>t) =
\int_0^\infty\int_0^\alpha
P_0(r,\theta,t\mid r_0,\theta_0)\,r\,d\theta\,dr.
\end{align}
Consequently, the first-passage-time density to any boundary is given by \cite{redner2001guide}
\begin{align}\label{fpt_intext}
    f_0(t)=-\frac{d Q_0(t)}{dt},
\end{align}
from which the moments can be defined directly. For instance, the mean first-passage time (MFPT) reads
\begin{align}
\langle T_0 \rangle=\int_0^\infty dt~ t f_0(t)=\int_0^\infty dt~ Q_0(t)
\label{MFPT-def}
\end{align}

It is quite useful to define the probability current in such set-up 
since it allows us to detect which boundary is reached by the particle, and the corresponding stochastic time. We therefore denote $J_i(t)$ the first-passage density associated with a first exit through boundary $i$ so that
\begin{align}
    J_i(t)
    &=
    \frac{d}{dt}
    \Pr\bigl(T_0\leq t,\ \theta(T_0)=i\bigr),
~~  i\in\{0,\alpha\}.
    \label{combined_flux_definitions}
\end{align}
In our set-up, these boundary currents at the wedge boundaries follow from Fick's law
\begin{align}\label{boundary_current_definitions}
J_0(t)
&=
D\int_0^\infty
\frac{1}{r}
\left.
\frac{\partial P_0(r,\theta,t\mid r_0,\theta_0)}{\partial\theta}
\right|_{\theta=0}
dr,
\nonumber\\
J_\alpha(t)
&=
-D\int_0^\infty
\frac{1}{r}
\left.
\frac{\partial P_0(r,\theta,t\mid r_0,\theta_0)}{\partial\theta}
\right|_{\theta=\alpha}
dr.
\end{align}
Since every first exit occurs through one of them, probability conservation gives
\begin{align}
    f_0(t)
    &=
    J_0(t)+J_\alpha(t)
    =
    -\frac{dQ_0(t)}{dt}.
    \label{tot_flux}
\end{align}

Finally, integrating each boundary contribution over all the exit time gives the splitting probabilities
\begin{align}\label{model_splitting_probabilities}
\epsilon_i
&=\int_0^\infty J_i(t)\,dt
=\Pr\bigl(\theta(T_0)=i\bigr),
~~i\in\{0,\alpha\},
\end{align}
which satisfy $\epsilon_0+\epsilon_\alpha=1$. Thus $J_i(t)/\epsilon_i$ is the exit-time density conditioned on reaching boundary $i$. Throughout the paper, $P_0,Q_0, J_i~\text{and}~\epsilon_i$ will refer to the non-reset process, while the fundamental relations connecting them should hold for reset process as well.

\section{\label{target}Target Search under Pure Threshold Resetting}

We first consider the target search solely under threshold resetting protocol. Our interest is to determine the statistics of the first-passage time of the particle to arrive at the absorbing
target at $\theta=0$. The second boundary at $\theta=\alpha$ acts as a threshold boundary
whose first encounter instantaneously resets the particle to its initial position
$\vec r_0=(r_0,\theta_0)$ and starts a new search cycle. Let $T_{TR}$ denote the total time to target absorption, including all preceding threshold resets.

\subsection{Mean first-passage time}
A natural observable is the mean first-passage time (MFPT) for the diffusing searcher to locate the absorbing target at $\theta=0$ for this configuration. This was derived by Biswas \textit{et al} in \cite{Biswas2025_Threshold} and it is expressed as a ratio of two boundary-integrated quantities from the underlying reset-free process, namely
\begin{equation} \label{general_threshold_mfpt}
  \langle T_{TR} \rangle = \frac{\int_0^\infty Q_{0}(t)\, dt}{\int_0^\infty J_0(t)\, dt},
\end{equation}
where $Q_0(t)$ and $J_0(t)$ are the non-reset quantities defined in Sec.~\ref{models}. Their
time integrals give $\langle T_0 \rangle$ [Eq. (\ref{MFPT-def})] and splitting probability [Eq. (\ref{model_splitting_probabilities})] respectively \cite{redner2001guide}.
Substituting these definitions yields a compact form for \eref{general_threshold_mfpt} resulting in
\begin{equation} \label{compact_threshold_mfpt}
  \langle T_{TR} \rangle = \frac{\langle T_0 \rangle}{\epsilon_0}.
\end{equation}
For simple diffusion, it is well known that the splitting probability satisfies the Laplace's equation \cite{redner2001guide}
\begin{equation}
    \nabla^{2}\epsilon_0 = 0,
\end{equation}
which in polar coordinates reads
\begin{equation}
    \frac{1}{r}\frac{\partial}{\partial r}
    \!\left(r\,\frac{\partial\epsilon_0}{\partial r}\right)
    +\frac{1}{r^{2}}\frac{\partial^{2}\epsilon_0}{\partial\theta^{2}}=0.
\end{equation}
The wedge domain which is a region bounded by two infinite rays meeting at the origin, contains no 
characteristic length scale. Rescaling $r\to\Gamma r$ for any $\Gamma>0$ 
leaves the domain, both boundaries, and the equation unchanged, so 
$\epsilon_0(r,\theta)=\epsilon_0(\Gamma r,\theta)$ for every $\Gamma>0$. 
The only functions with this property are those independent of $r$ entirely. Therefore the radial term vanishes and the 
equation reduces to the elementary ordinary differential equation(ODE),
\begin{equation}
    \frac{d^{2}\epsilon_0}{d\theta^{2}}=0,
\end{equation}
whose general solution is linear, $\epsilon_0(\theta)=A\theta+B$. Imposing 
the boundary conditions $\epsilon_0(0)=1$ (certain capture at the target) 
and $\epsilon_0(\alpha)=0$ (no capture upon reaching the threshold) gives 
$A=-1/\alpha$ and $B=1$, so
\begin{equation}
    \epsilon_0(\theta) = 1 - \frac{\theta}{\alpha}.
\end{equation}
For a particle starting at $\theta_{0}=u\alpha$ the splitting probabilities are 
therefore
\begin{equation}\label{split_u}
    \epsilon_0 = 1 - u,\qquad \epsilon_\alpha=u
\end{equation}
a result that depends only on the fractional angular position $u$ and is 
independent of the radial distance from the origin~\cite{redner2001guide}.

\begin{figure}[htbp]
  \centering \includegraphics[width=\columnwidth]{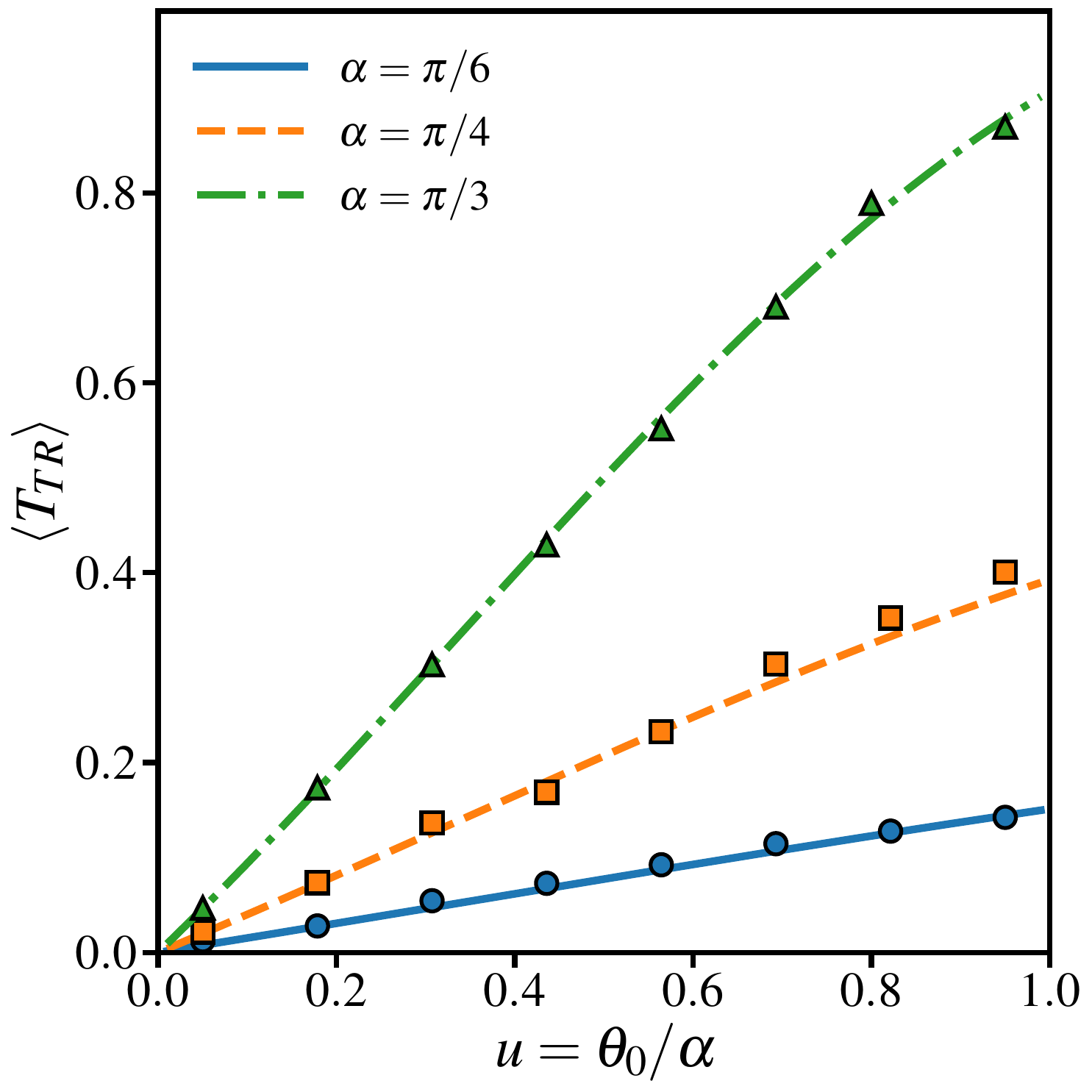}
  \caption{The analytical prediction ~\eref{exact_T_TR} of the MFPT under threshold resetting (lines) is compared against Langevin dynamics simulations for wedge angles $\alpha=\pi/6$ (circles), $\pi/4$ (squares), $\pi/3$ (triangles). Langevin dynamics simulations are performed for $N=3\times10^4$ independent searches with timestep $\Delta t = 10^{-7}$ and $r_0 = 1.0$, $D = 1.0$.}
  \label{single_mfpt_verification}
\end{figure}

Substituting this splitting probability to target along with the well known expression for the unconditional MFPT $\langle T_0 \rangle$ from \cite{redner2001guide,3bc4-nlkt} into Eq.~(\ref{compact_threshold_mfpt}), MFPT for the TR protocol becomes
\begin{equation} \label{exact_T_TR}
  \langle T_{TR} \rangle = \frac{r_0^2}{4D(1-u)} \left[ \frac{\cos(\alpha(2u - 1))}{\cos(\alpha)} - 1 \right],
  \qquad 0<\alpha<\frac{\pi}{2}.
\end{equation}
where $u = \theta_0/\alpha \in (0,1)$ is the dimensionless target distance. In \fref{single_mfpt_verification}, we plot \eref{exact_T_TR} for different apex angles $\alpha$ along with the same quantity obtained from direct simulation of the model (See Appendix \ref{numerical_simulations} for the details regarding simulations).

Analyzing the expression~\eref{exact_T_TR} reveals two kinetic features of the pure threshold protocol. First, evaluating the derivative with respect to the initial configuration $u$ confirms that the search time is strictly monotonically increasing (\fref{single_mfpt_verification}). Consequently, there is no internal optimal initiation point $u^*$ that minimizes the MFPT; the search time trivially minimizes as the initiation point approaches the target. 
Second, in the limit of extreme proximity to the  target ($\theta_0 \to 0$), performing a lowest-order Taylor expansion yields the precise asymptotic behavior:
\begin{equation}
  \langle T_{TR} \rangle \simeq \frac{r_0^2 \tan(\alpha)}{2D} \theta_0 \quad \text{as } \theta_0 \to 0,
\end{equation}
which guarantees $\langle T_{TR}\rangle \to0$ as the searcher initiates closer and closer to the target$(u\to0)$.
Writing the initial arc-length distance to the target as $s_0=r_0\theta_0$, this result becomes $\langle T_{TR}\rangle\simeq s_0L_\text{eff}/(2D)$, with $L_\text{eff}=r_0\tan\alpha$. This leading-order approximation is analogous to the near-boundary result $T(x)\simeq x_0L/(2D)$ for diffusion in a finite interval with absorbing endpoints~\cite{redner2001guide}. Here, the effective geometric length $L_\text{eff}$ grows with the wedge angle and diverges as $\alpha\to\pi/2$.

For a fixed wedge angle in this finite-mean domain, the one-sided limit at
the threshold is finite 
\begin{align}\label{threshold_mfpt_u_to_one}
\lim_{u\to1^-}\left\langle T_{TR}\right\rangle
=
\frac{r_0^2\alpha\tan\alpha}{2D}.
\end{align}
Although the limiting value is finite, the endpoint $u=1$ is excluded
because placing the reset point directly on the threshold would produce an
immediate-reset repetition.

\subsection{Second moment of first-passage time}
\label{sec:second_moment}

Using the exact renewal formalism (See Appendix~\ref{app:renewal_moments}), the moments of the threshold-resetting process can be constructed directly from the boundary-resolved moments of the non-reset process,
\begin{align}\label{partial_moment_definition}
m_{n,i}&=\int_0^\infty t^nJ_i(t)\,dt,
\qquad i\in\{0,\alpha\},
\end{align}
whenever these moments are finite. The corresponding moment conditioned on reaching boundary $i$ is $m_{n,i}/\epsilon_i$, and
\begin{align}\label{partial_moment_sum}
m_{n,0}+m_{n,\alpha}=\langle T_0^n\rangle.
\end{align}
where $\langle T_0^n\rangle$ is the unconditional $n^{th}$ moment of the non-reset first-exit time defined in Eq.~\eqref{combined_spatial_exit_timeNEW}. 
For TR, the second moment is obtained as(See Appendix~\ref{app:renewal_moments}),
\begin{align}
    \left\langle T_{TR}^2\right\rangle
=
\frac{\left\langle T_0^2\right\rangle}{\epsilon_0}
+
\frac{
2m_{1,\alpha}\left\langle T_0\right\rangle
}{
\epsilon_0^2
}
\label{second_moment_unevaluated}
\end{align}
where $\epsilon_0$ is given by Eq.~\eqref{split_u}, and $\langle T_0\rangle , \langle T_0^2\rangle $ are the first two moments of the non-reset exit times defined in Eq.~\eqref{combined_spatial_exit_timeNEW}.  

After evaluating $m_{1,\alpha}$ and substituting the exact moments of the free-wedge process in Eq.~\eqref{second_moment_unevaluated}, yields the following closed-form expression for the second moment of the threshold-resetting time,
\begin{align}
\left\langle T_{TR}^2\right\rangle
&=\frac{r_0^4}{96D^2(1-u)}
\Biggl[\frac{\cos\!\left(2\alpha(1-2u)\right)}{\cos(2\alpha)}
-\frac{4\cos\!\left(\alpha(1-2u)\right)}{\cos\alpha}
+3\Biggr]
\nonumber\\
&+\frac{r_0^4}{8D^2(1-u)^2}
\left[\frac{\cos\!\left(\alpha(1-2u)\right)}{\cos\alpha}-1\right]
\times\left[\frac{\sin(2\alpha u)}{\sin(2\alpha)}-u\right],~~ \nonumber \\
& ~~ \left(0<\alpha<\frac{\pi}{4}\right).
\label{second_moment_full_expression}
\end{align}
The divergence of the fluctuations is immediately apparent in the denominator of the first term. The presence of $\cos 2\alpha$ dictates that the second moment, and therefore the variance, strictly diverges as the wedge angle approaches $\alpha \to \pi/4$. This can be explicitly confirmed under the asymptotic analysis of the first passage time density, which we present in the following section~\ref{tail_analysis}.

\subsection{Asymptotic analysis of the first-passage-time density}
\label{tail_analysis}
We now examine the long-time behavior of the first-passage-time density
under threshold resetting. The particle diffuses in the wedge
$0<\theta<\alpha$, while the radial coordinate remains unbounded,
$r>0$. Consequently, trajectories can undergo arbitrarily long radial
excursions before reaching either angular boundary.

\begin{figure}[htbp]
    \centering
    \includegraphics[width=1\linewidth]{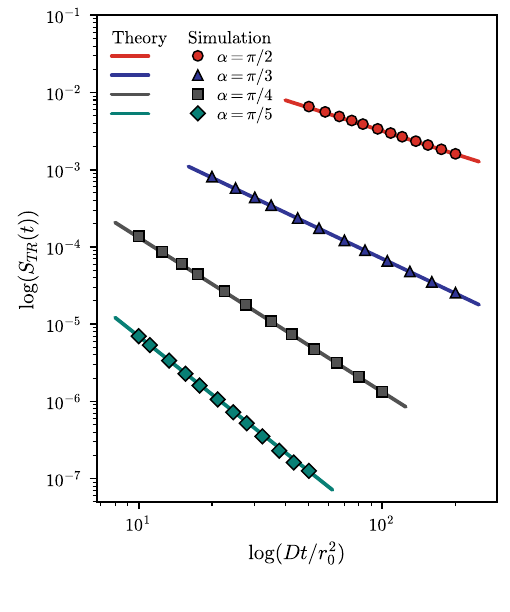}
    \caption{The large time behavior of the survival probability under threshold resetting plotted against the dimensionless time $Dt/r_0^2$ on logarithmic scale (Solid lines). Symbols show Brownian motion simulation data for wedge angles $\alpha=\pi/2$ (circles), $\pi/3$ (triangles), $\pi/4$ (squares), and $\pi/5$ (diamonds) and the solid lines fit a power-law decay $t^{-\pi/(2\alpha)}$ as in \eref{main_dimensionless_survival_tail}. Each dataset contains $5\times10^8$ independent searches, with $u=\theta_0/\alpha=1/2$, $r_0=D=1$, and no Poissonian resetting.}
    \label{long_time_tail_verification}
\end{figure}

For the non-reset process, where both boundaries act as absorbing targets, the survival probability has the long time form \cite{majumder2015}, \begin{align}
\label{long_time_survival}
Q_0(t)\sim A_0t^{-\beta},
\end{align}
where $\beta=\pi/(2\alpha)$ and $A_0>0$. For the threshold-reset completion time $T_{TR}$, we write 
\begin{align}\label{threshold_survival_density_definition}
S_{TR}(t)&=\Pr(T_{TR}>t),
\nonumber\\
f_{TR}(t)&=-\frac{dS_{TR}(t)}{dt}.
\end{align}
Survival up to time $t$ can occur either without encountering the either boundary, or after the first cycle ends at the threshold at time $\tau<t$ and the renewed search survives for the remaining time $t-\tau$. Using the boundary current defined in Eq.~\eqref{combined_flux_definitions}, we can write the following renewal equation \cite{Biswas2025_Threshold,biswas2026optimal} 
\begin{align}
    S_{TR}(t)
    =
    Q_0(t)
    +
    \int_0^t
    J_\alpha(\tau)S_{TR}(t-\tau)\,d\tau.
    \label{first_renewal_survival_intext}
\end{align}
For a time-dependent function $h(t)$, we use the Laplace transform notation
\begin{align}\label{laplace_transform_convention}
\widetilde h(s)=\int_0^\infty e^{-st}h(t)\,dt,
\qquad s>0.
\end{align}
Under this notation, the transforms of $Q_0(t)$ and $J_i(t)$ are denoted by $\widetilde q_0(s)$ and $\widetilde j_i(s)$, respectively. Transforming the renewal relation gives
\begin{align}
    \widetilde S_{TR}(s)
    =
    \frac{\widetilde q_0(s)}
    {1-\widetilde j_\alpha(s)}.
    \label{renewal_laplace_intext}
\end{align}
Since $S_{TR}(0)=1$, Eq.~\eqref{threshold_survival_density_definition} also gives $\widetilde f_{TR}=1-s\widetilde S_{TR}(s)$. Using the small-$s$ behavior of the non-reset transforms in this renewal relation determines leading non-analytic contributions to the first passage density $\widetilde f_{TR}(s)$ (see Appendix~\ref{app:laplace_branch_cut}),
\begin{align}
    &\widetilde f_{TR}(s)
    =
    1+\cdots
    \nonumber\\
    &\quad+
    \begin{cases}
        \mathcal A_1s^\beta+o(s^\beta),
        & \beta\notin\mathbb N,
        \\[2mm]
        \begin{aligned}
            &\mathcal A_2s^\beta\log\left(\frac{r_0^2}{D}s\right)
         +o\!\left(s^\beta|\log\left(\frac{r_0^2}{D}s\right)|\right),
        \end{aligned}
        & \beta\in\mathbb N.
    \end{cases}
    \label{main_threshold_small_s}
\end{align}
The dot $\dots$ denote integer powers of $s^k$ with $1\le k< \beta$, whose coefficients are determined by the finite moments of the first passage time density. The coefficients $\mathcal A_1~\text{and}~\mathcal A_2$ include the geometric factors Eqs.~\eqref{app_density_coefficient_noninteger} and \eqref{app_density_coefficient_integer}. Standard inverse-Laplace asymptotics yield the following large time behavior (Appendix~\ref{app:laplace_branch_cut})
\begin{align}\label{long_tails_of_Sandf}
    S_{TR}(t)&\sim\frac{\mathcal C}{\beta}t^{-\beta}\quad &t\to\infty,\\
    f_{TR}(t)&\sim \mathcal C\,t^{-(1+\beta)}\quad &t\to\infty,
\end{align}
where $\mathcal C=\frac{\beta A_0}{1-u}>0$ is the corresponding tail amplitude. Threshold resetting therefore does not change the first-passage-time tail into an exponential
cutoff unlike Poissonian resetting; the algebraic tail originates from the unbounded radial direction.
Figure~\ref{long_time_tail_verification} compares this prediction
with Brownian simulations. Writing $t_D=r_0^2/D$, on the dimensionless time axis $t/t_D=Dt/r_0^2$, the theoretical curve is the logarithm of
\begin{align}
    S_{TR}(t)
    &\sim
    \frac{A_0}{(1-u)t_D^\beta}
    \left(\frac{t}{t_D}\right)^{-\frac{\pi}{2\alpha}}.
    \label{main_dimensionless_survival_tail}
\end{align}

\section{Target Search under the Combined Reset Protocol}
\label{combined_fpt}

In addition to the threshold-triggered reset, we let the diffusing particle also undergo Poissonian resetting at a rate $\lambda$
to the initial position
$\vec r_0=(r_0,\theta_0)$ inside the wedge (Fig.~\ref{fig:combined_schematic}). Now the trajectory of the diffusive searcher is
interrupted by whichever of three distinct physical events occurs
first: striking the absorbing target, which terminates the search, or encountering either the threshold boundary ($\theta=\ \alpha$) or the activation of the the stochastic reset, both of which instantly restart the search from
$\vec r_0$. Under this combined reset protocol, we derive an expression for the mean first-passage time to hit the absorbing target boundary at $\theta=0$.

Recently, García-Valladares \textit{et al.}~\cite{garcia2026target} studied the combined action of two resetting mechanisms for a Brownian particle by extending the diffusion equation with stochastic resetting, originally introduced by Evans and Majumdar~\cite{evans2011diffusion,evans2011prl}. While this approach provides a natural description for the specific diffusive dynamics considered therein, a more general framework is desirable to accommodate the interplay of threshold and Poissonian resetting for arbitrary underlying dynamics, including both Markovian and non-Markovian processes, and in different geometries. In this work, we develop such a framework by extending the random-variable-based renewal formalism introduced respectively in~\cite{pal2017first} for stochastic resetting and in \cite{biswas2026optimal} for TR to combine both the resetting mechanisms. Unlike approaches relying on an explicit evolution equation for a particular dynamics as in \cite{garcia2026target}, this formulation is built upon the renewal structure of the  random times involved and thus provides a unified route to studying their combined effects across a broader class of search processes. We illustrate this formalism in the following.

\subsection{Renewal formalism and the combined mean first-passage time}
\label{combined_mfpt_derivation}

Let $T_{\mathrm{tot}}$ denote the time to reach the target under
threshold resetting and an additional independent resetting mechanism.
We use the non-reset exit time $T_0$ defined in
Eq.~\eqref{combined_spatial_exit_timeNEW}, and let $R$ denote the
waiting time for the stochastic resets triggered by an external timer. At this stage, we assume the
distribution of $R$ to be general (and not necessarily exponential as in PR). After either reset, the particle
returns to $\vec r_0$ and a fresh waiting time is drawn independently
from the same distribution. Following the first-cycle renewal
approach~\cite{pal2017first},
\begin{equation}
    T_{\mathrm{tot}}=
    \begin{cases}
        T_0,
        & T_0<R,\ \theta(T_0)=0,\\[1mm]
        T_0+T_{\mathrm{tot}}',
        & T_0<R,\ \theta(T_0)=\alpha,\\[1mm]
        R+T_{\mathrm{tot}}'',
        & R<T_0
    \end{cases}.
    \label{combined_time_recursion}
\end{equation}
In the first case, the particle reaches the target before the
stochastic reset, ending the search. In the second, it reaches the
threshold first and starts a renewed search. In the third, the
stochastic reset occurs before either boundary is reached and the process restarts. The
remaining search times $T_{\mathrm{tot}}'$ and $T_{\mathrm{tot}}''$ in the second and third case respectively
are statistically independent copies of $T_{\mathrm{tot}}$ i.e., they are independent of each other and the
first cycle time. Thus, Eq. (\ref{combined_time_recursion}) suggests three mutually exclusive events which occur with the following probabilities 
\begin{align}
P_\text{abs}&=\Pr\!\left(T_0<R,\ \theta(T_0)=0\right),\nonumber\\
    P_T&=\Pr\!\left(T_0<R,\ \theta(T_0)=\alpha\right),
    \nonumber \\
    P_p&=\Pr(R<T_0),
    \label{combined_cycle_probabilities}
\end{align}
and they conserve probability
\begin{align}
    P_\text{abs}+P_T+P_p=1.
    \label{combined_probability_normalization}
\end{align}
Physically, $P_{\mathrm{abs}}$ denotes the probability that the searcher reaches the absorbing boundary at $\theta=0$ before a Poissonian reset occurs, $P_T$ denotes the probability that the searcher reaches the threshold boundary at $\theta=\alpha$ before a Poissonian reset occurs, and $P_p$ denotes the probability that a Poissonian reset occurs before the searcher reaches either boundary.

The renewal equation (\ref{combined_time_recursion}) can, equivalently, be written as
\begin{align}
    T_{\mathrm{tot}}
    =\min(T_0,R)
    +I\!\left(T_0<R,\ \theta(T_0)=\alpha\right)
    T_{\mathrm{tot}}'
    +I(R<T_0)T_{\mathrm{tot}}'',
    \label{TCR}
\end{align}
where we have defined indicator function as follows
\begin{equation}
    I(E)=
    \begin{cases}
        1, & \text{if $E$ occurs},\\
        0, & \text{otherwise}.
    \end{cases}
    \label{combined_indicator_definition}
\end{equation}
Applying the law of total expectations on both sides of Eq. (\ref{TCR}), we find 
\begin{align}
    \langle T_{\mathrm{tot}}\rangle
    &=\langle\min(T_0,R)\rangle
    +\left\langle
    I\!\left(T_0<R,\ \theta(T_0)=\alpha\right)\right\rangle
  \langle   T_{\mathrm{tot}}' \rangle
    \nonumber\\
    &+\left\langle
    I(R<T_0)\right\rangle \langle T_{\mathrm{tot}}'' \rangle,
    \label{TCR-exp}
\end{align}
where the expectations of the indicators are essentially the probabilities defined in Eq. (\ref{combined_cycle_probabilities}) namely
\begin{align}
   P_T &=\left\langle
    I\!\left(T_0<R,\ \theta(T_0)=\alpha\right)\right\rangle \nonumber\\
   P_p &=\left\langle
    I(R<T_0)\right\rangle 
\end{align}
Simplifying Eq. (\ref{TCR-exp}) and using the above-defined probabilities, we arrive at the exact expression for the combined MFPT
\begin{equation}
    \left\langle T_{\mathrm{tot}}\right\rangle
    =\frac{\left\langle\min(T_0,R)\right\rangle}{P_\mathrm{abs}}.
    \label{combined_mfpt_cycle_form}
\end{equation}
Thus, the mean search time is the mean duration of a cycle divided
by its probability of ending at the target. This relation holds
for any distribution of $R$ under the independent renewal
assumptions above, provided $P_\mathrm{abs}>0$. The MFPT is finite when
$\langle\min(T_0,R)\rangle$ is finite.

 We now specialize to Poissonian resetting with rate $\lambda>0$ so that
\begin{equation}
    f_R(t)=\lambda e^{-\lambda t},
    \qquad
    \Pr(R>t)=e^{-\lambda t},
    \quad t\geq0.
    \label{combined_poisson_density}
\end{equation}
On the other hand 
\begin{align}
    \Pr(\min(T_0,R)>t)=\Pr(T_0>t) \Pr(R>t)=Q_0(t)e^{-\lambda t},
\end{align} and hence
\begin{align}
    \left\langle\min(T_0,R)\right\rangle
    &=\int_0^\infty Q_0(t)e^{-\lambda t}\,dt
    =\widetilde q_0(\lambda),
    \label{combined_mean_cycle_duration}
\end{align}
using the Laplace-transform convention in
Eq.~\eqref{laplace_transform_convention}.
We can further simplify the probabilities in terms of the non-reset boundary currents defined in
Eq.~\eqref{combined_flux_definitions}. In particular, the probabilities $P_{\mathrm{abs}}$ and $P_T$ of reaching the absorbing and threshold boundaries, respectively, before a Poissonian reset are obtained by weighting the corresponding boundary currents by the probability $e^{-\lambda t}$ of avoiding Poissonian resetting up to time $t$. On the other hand, the probability $P_p$ of a Poissonian reset occurring before either boundary is reached is obtained by weighting the survival probability $Q_0(t)$ by the Poissonian reset-time density $f_R(t)=\lambda e^{-\lambda t}$. This leads to the following expressions
\begin{subequations}
\label{cycle_probabilities}
\begin{align}
     P_\mathrm{abs}
    &=\int_0^\infty J_0(t)e^{-\lambda t}\,dt
    =\widetilde j_0(\lambda),
    \label{combined_absorption_probability}\\
    P_T
    &=\int_0^\infty J_\alpha(t)e^{-\lambda t}\,dt
    =\widetilde j_\alpha(\lambda),
    \label{combined_threshold_probability}\\
    P_p
    &=\int_0^\infty \lambda e^{-\lambda t}Q_0(t)\,dt
    =\lambda\widetilde q_0(\lambda).
    \label{combined_poissonian_probability}
\end{align}
\end{subequations}
Substitution Eqs. (\ref{combined_mean_cycle_duration}) and (\ref{combined_absorption_probability}) into Eq.~\eqref{combined_mfpt_cycle_form} yields an expression for the combined MFPT
\begin{equation}
    \left\langle T_{\mathrm{tot}}\right\rangle
    =\frac{\widetilde q_0(\lambda)}
    {\widetilde j_0(\lambda)},
    \label{combined_mfpt_derived}
\end{equation}
that can be evaluated from the underlying reset-free process, which is in the same spirit as the original resetting problems.

\begin{figure}[H]
    \centering
    \includegraphics[width=\linewidth]{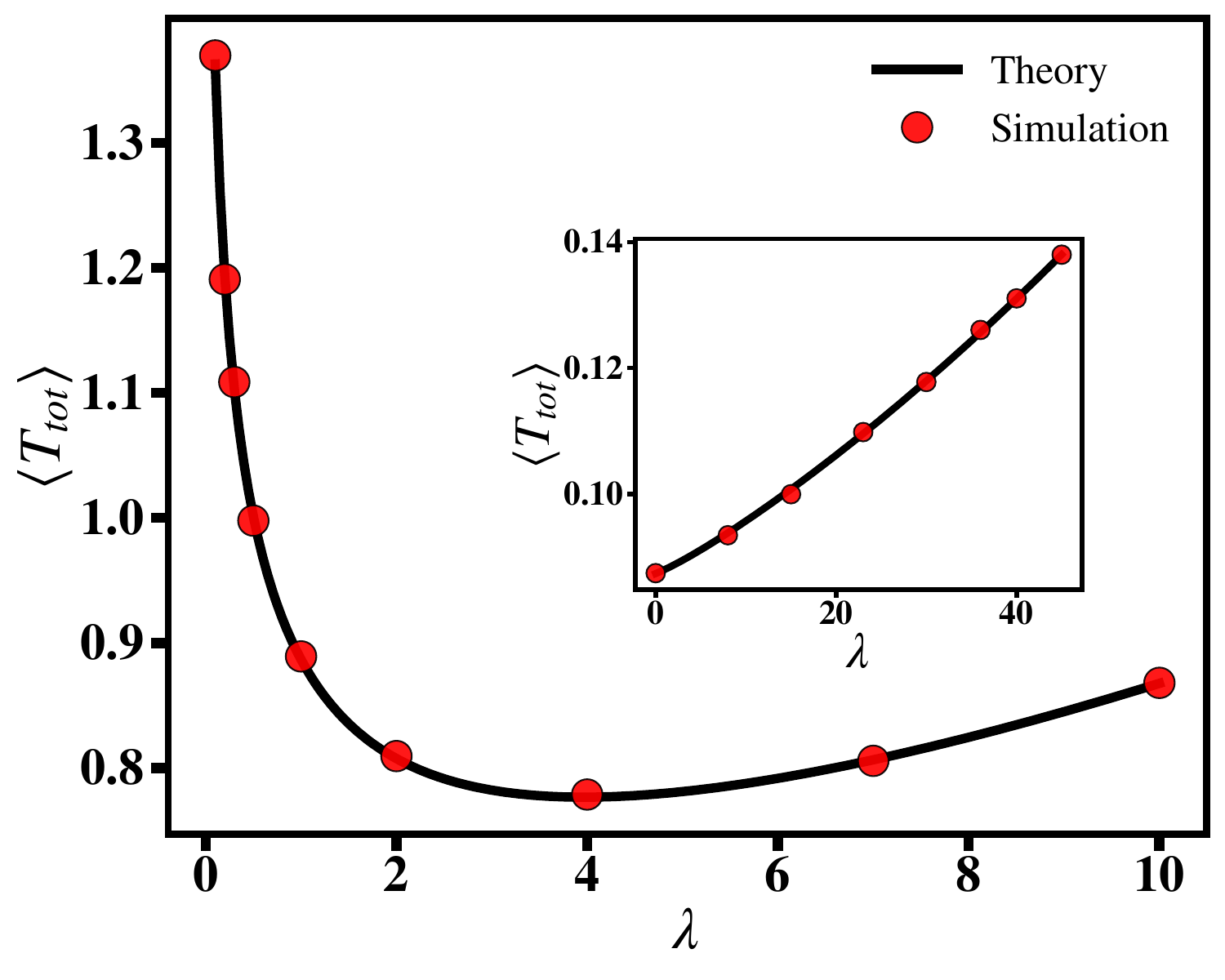}
    \caption{Numerical validation of the mean first-passage time $\langle T_{\mathrm{tot}}\rangle$ under combined threshold and Poissonian resetting. Analytical predictions [\eref{combined_mfpt_derived}] depicted as solid lines are compared with Langevin dynamics simulations (circles). The main panel corresponds to a wedge with $\alpha=\pi/2$ and $u=0.5$. The inset shows the corresponding results for $\alpha=\pi/7$ and $u=0.8$.
    For both configurations, $r_0=D=1$.}
    \label{fig:universal_verification}
\end{figure}

We verify Eq.~\eqref{combined_mfpt_derived} in Fig.~\ref{fig:universal_verification} for different wedge angles. Interestingly, we find that the effect of Poissonian resetting is not universal: it reduces the mean completion time for $\alpha=\pi/2$, whereas it increases it for $\alpha=\pi/7$. This contrasting behaviour highlights that introducing an additional resetting mechanism does not necessarily improve an already existing threshold-resetting process. Rather, its effectiveness depends sensitively on the underlying search dynamics and the geometric parameters of the system. This raises a natural question: what determines whether Poissonian resetting enhances or hinders a threshold-resetting search? In the following, we address this question by developing a criterion that elucidates the role of the system parameters in governing the competition between the two resetting mechanisms.

\subsection{Coefficient of variation and the resetting criterion}
\label{sec:simplified_cv}
The mean target search time for diffusive search with TR, when combined with Poissonian resetting can be affected both favorably and adversely, which may be seen in \fref{fig:universal_verification}, where the MFPT for the combined protocol $\langle T_{tot}\rangle$ shows a reduction as a function of resetting rate $\lambda$ in one set of system parameters and a monotonic increase for another set. It would therefore be useful to have a phase diagram in terms of the relevant parameters which are the initial scaled angle $u$ and the wedge apex angle $\alpha$, where one can demarcate when resetting is beneficial and otherwise. We note here that  the radial coordinate gets scaled out in the expression for the MFPT, keeping the relevant variables to $u$ and $\alpha$.

\begin{figure}[tb]
    \centering
    \includegraphics[width=\columnwidth]{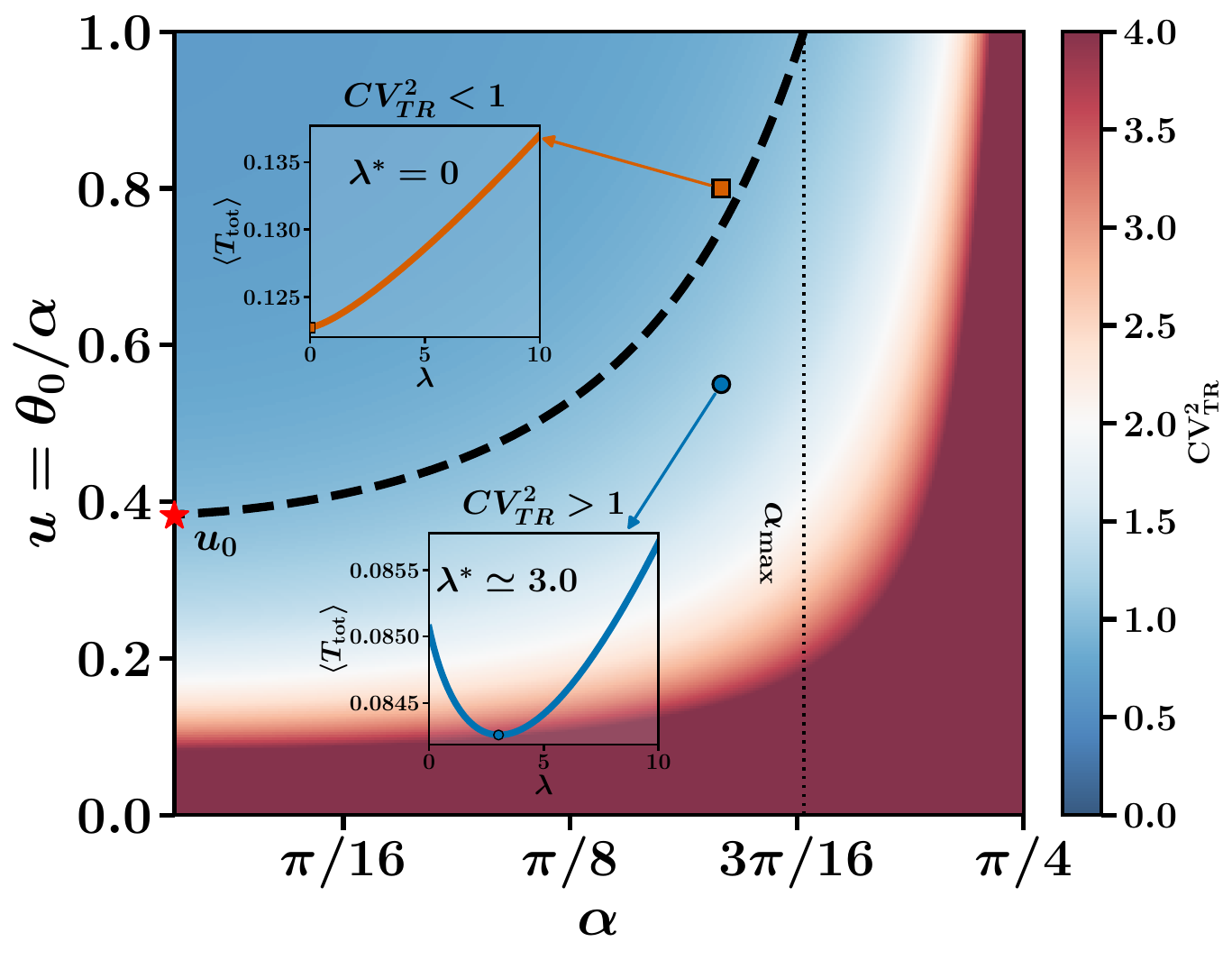}
    \caption{The squared coefficient of variation, $CV_{TR}^2$ [See \eref{cv_definition}] mapped across the parameter space of the wedge angle $\alpha$ and the dimensionless initial position $u = \theta_0/\alpha$. The dashed contour delineates the crossover ($CV_{TR}^2 = 1$) between the parameter regions where Poissonian resetting is beneficial in reducing the MFPT ($CV_{TR}^2>1$) and the one where it is not ($CV_{TR}^2<1$). The red star on the $u$ axis marks $u_0=(3-\sqrt{5})/2$, the limiting position of this boundary as $\alpha\to0^+$. The vertical black dotted line marks $\alpha_{\max}\simeq0.59504~\mathrm{rad}$, the limiting wedge angle of the boundary as $u\to1^-$. The blue circle at $(\alpha,u)=(\pi/6,0.55)$ and the orange square at $(\pi/6,0.8)$ lie in the $CV_{TR}^2>1$ and $CV_{TR}^2<1$ regions, respectively. The corresponding insets show the mean first-passage time $\langle T_{tot}\rangle$ under combined threshold and Poisson resetting as a function of the Poissonian reset rate $\lambda$, with $r_0=D=1$. The blue curve has a minimum at $\lambda>0$, whereas the orange curve has its minimum at $\lambda=0$.}
    \label{fig:cv2_landscape}
\end{figure}

A formal approach to realize this has been considered earlier \cite{reuveni2016optimal,pal2017first,ahmad2019first,pal2019landau} which is to analyze the $\lambda\to 0$ behavior of the coefficient of variation  of the underlying search problem problem, i.e.,  in the absence of Poissonian resetting. In the present case where the underlying process is the diffusion under TR, the coefficient of variation $CV_{TR}$ is given by,
\begin{align}\label{cv_definition}
CV_{TR}
=
\frac{\sqrt{\left\langle T_{TR}^2\right\rangle
-
\left\langle T_{TR}\right\rangle^2}}{\left\langle T_{TR}\right\rangle}.
\end{align}
If $CV_{TR}^2>1$, the derivative  $d\langle T_{tot}\rangle/d\lambda<0$  in the limit $\lambda\to 0$,  leading to a reduction of MFPT $\langle T_{tot}\rangle$ for  $\lambda>0$. Conversely when $CV_{TR}^2>1$, $d\langle T_{tot}\rangle/d\lambda>0$ the limit $\lambda\to 0$, indicating enhancement  in $\langle T_{tot}\rangle$ for  $\lambda>0$.
Hence, the condition $CV_{TR}^2(u,\alpha)=1$ 
marks the boundary between these two behaviors in the $(u,\alpha)$ phase diagram. Such a phase diagram is displayed in \fref{fig:cv2_landscape}.

In  \fref{fig:cv2_landscape}, we plot $CV_{TR}^2(u,\alpha)$ as a function of $u$ and $\alpha$, calculated from \eref{cv_definition} using the known expressions for the first and second moments of the underlying process under threshold resetting,  $\langle T_{TR}\rangle$ [\eref{exact_T_TR}] and  $\langle T^2_{TR}\rangle$ [\eref{second_moment_full_expression}], expressed as (See Appendix.\ref{app:threshold_cv} for details),
\begin{align}
CV_{TR}^2(u,\alpha)
=
\frac{1-u}{6}
\frac{B_2(u,\alpha)}{B_1^2(u,\alpha)}
+
2\frac{B_3(u,\alpha)}{B_1(u,\alpha)}
-1.
\label{simplified_cv}
\end{align}
Here,
\begin{subequations}
\label{B_functions_definition}
\begin{align}
B_1(u,\alpha)
&=
\frac{\cos\!\left[\alpha(1-2u)\right]}{\cos\alpha}-1,
\label{b1}
\\
B_2(u,\alpha)
&=
\frac{\cos\!\left[2\alpha(1-2u)\right]}{\cos(2\alpha)}
-[4B_1(u,\alpha)+1],\label{b2}
\\
B_3(u,\alpha)
&=
\frac{\sin(2\alpha u)}{\sin(2\alpha)}-u\label{b3}.
\end{align}
\end{subequations}
In the \fref{fig:cv2_landscape}, the dashed line - obtained by solving  \eref{simplified_cv} numerically - demarcates the cross over between regions where resetting serves to slow down the typical time to reach the target (region on top left) and the one where the resetting allows to expedite the search time. In the plot, even though we have limited the range of wedge angles to $0<\alpha<\pi/4$, the conclusion that resetting is helpful for arbitrary initial angles for $\alpha>\pi/4$ holds true, as $CV_{TR}^2$ diverges in this region.

As shown in Fig.~\ref{fig:cv2_landscape}, the end point of the separatrix line $CV_{TR}=1$, when $\alpha\to0$ (highlighted by a star symbol in color red) indicates a minimum  scaled angle of resetting ($u_0$) below which, the diffusive search with TR benefits in reducing the MFPT by the introduction of Poissonian resetting. 
In this limit
\begin{align}
\lim_{\alpha\to0^+}CV_{TR}^2(u,\alpha)
=
\frac{1+u^2}{3u},
\end{align}
so the condition $CV_{TR}=1$ reaches $\alpha=0$ at
\begin{align}
u_0=\frac{3-\sqrt5}{2}.
\end{align}

Further, when $u\to1$ along the crossover line $CV_{TR}=1$ in \fref{fig:cv2_landscape} denotes the   the maximum wedge angle $\alpha_{max}$ 
beyond which, the introduction of Poissonian resetting guarantees the reduction of MFPT.
To obtain $\alpha_{max}$, we take the limit $u\to1$ from below, in the expression for $CV_{TR}^2(u,\alpha)$ [\eref{simplified_cv}] leading to 
\begin{align}
\lim_{u\to1^-}CV_{TR}^2(u,\alpha)
&=
\frac{1}{\alpha\tan\alpha}
-\frac{1}{\tan^2\alpha}
+\nonumber\\&\frac{\tan\alpha}{3\alpha[1-\tan^2\alpha]}.
\label{u-1limit-critical line}
\end{align}
Equating \eref{u-1limit-critical line} to unity results in,
\begin{align}
3\tan\alpha_{max}-2\tan^3\alpha_{max}
-3\alpha_{max}(1-\tan^4\alpha_{max})=0,
\end{align}
whose nonzero root $\alpha_{max}$ in $(0,\pi/4)$ is close to $\pi/16$, denoted by an indicative dotted line in \fref{fig:cv2_landscape}. 
In the inset of \fref{fig:cv2_landscape}, we verify the above claims by plotting the MFPT for combined protocol $\langle T_{tot}\rangle$, for two points in the $(u,\alpha)$ each taken from different domains, one corresponding to $CV_{TR}^2<1$ and the other to $CV_{TR}^2>1$.


\section{Reset Count Statistics and Correlations}
\label{sec:reset_counts}

The mean first-passage time measures the duration of the search,
but does not specify how many resetting interventions occur before
target absorption. Although threshold and Poissonian resetting
compete to end a cycle, a search may undergo many resets of both
types before reaching the target. We therefore examine their
individual and joint count statistics to determine how the two
counts are correlated and how this dependence changes with the
system parameters.

\subsection{Renewal description}
We use the renewal cycles introduced in
Sec.~\ref{combined_mfpt_derivation}, with outcome probabilities
$P_\mathrm{abs}$, $P_T$, and $P_p$ defined in
Eq.~\eqref{combined_cycle_probabilities}.
Let $\mathcal N_{th}$ and $\mathcal N_p$ denote the total numbers
of threshold and Poissonian resets, respectively, before target
absorption. 
These numbers collectively satisfy the following renewal equation
\begin{align}
    (\mathcal N_{th},\mathcal N_p) =
    \begin{cases}
        (0,0),
        & T_0<R,\ \theta(T_0)=0,
        \\[1mm]
        (1+\mathcal N_{th}',\mathcal N_p'),
        & T_0<R,\ \theta(T_0)=\alpha,
        \\[1mm]
        (\mathcal N_{th}'',1+\mathcal N_p''),
        & R<T_0.
    \end{cases}
    \label{joint_count_recursion}
\end{align}
Eq. (\ref{joint_count_recursion}) can be reasoned in the following way. 
Target absorption ends the search without a reset, so $(\mathcal N_{th},\mathcal N_p)=(0,0)$. A threshold
encounter adds unity to $\mathcal N_{th}$, whereas a Poissonian
reset adds one to $\mathcal N_p$; either reset then starts a
renewed search following which, the counts are renewed, denoted by the primes.
The primed pair has the same joint distribution as
$(\mathcal N_{th},\mathcal N_p)$ and is independent of the first
cycle. No independence between the two counts themselves is
assumed.

\subsection{Joint count mass function and averages}
\label{reset_distribution}
To compute the joint mass function of the counts, 
we first define the joint probability generating function
\begin{align}
    \mathcal G(z_{th},z_p)
    &=\left\langle z_{th}^{\mathcal N_{th}}
    z_p^{\mathcal N_p}\right\rangle \nonumber\\
    &=\sum_{m,n\geq0} \Pr(\mathcal N_{th}=m,\mathcal N_p=n) z_{th}^mz_p^n
    \label{count_generating_function_definition}
\end{align}
for $0\leq z_{th},z_p\leq1$.
Taking the expectation over the three outcomes in
Eq.~\eqref{joint_count_recursion} gives
\begin{align}
    \mathcal G(z_{th},z_p)
    &=P_\mathrm{abs} \left\langle z_{th}^{0}
    z_p^{0}\right\rangle
    + P_T \left\langle z_{th}^{1+\mathcal N_{th}'}
    z_p^{\mathcal N_{p}'}\right\rangle+P_p \left\langle z_{th}^{\mathcal N_{th}''}
    z_p^{1+\mathcal N_{p}''}\right\rangle \nonumber \\
    &=P_\mathrm{abs}+(P_Tz_{th}+P_pz_p)\mathcal G(z_{th},z_p).
    \label{count_generating_function_renewal}
\end{align}
Therefore,
\begin{align}
    \mathcal G(z_{th},z_p)
    =\frac{P_\mathrm{abs}}{1-P_Tz_{th}-P_pz_p}.
    \label{count_generating_function}
\end{align}
Expanding the denominator as a geometric series and then applying
the binomial expansion yields
\begin{align}
    \mathcal G(z_{th},z_p)
    &=P_\mathrm{abs}\sum_{k=0}^{\infty}
    (P_Tz_{th}+P_pz_p)^k
    \nonumber\\
    &=\sum_{m,n\geq0}\binom{m+n}{m}
    P_T^mP_p^nP_\mathrm{abs}z_{th}^mz_p^n.
    \label{count_generating_function_expansion}
\end{align}
Identifying the coefficients with Eq. (\ref{count_generating_function_definition}) gives the joint count mass function
\begin{align}
    \Pr(\mathcal N_{th}=m,\mathcal N_p=n)
    =\binom{m+n}{m}P_T^mP_p^nP_\mathrm{abs},
    ~ [m,n=0,1,2,\ldots].
    \label{main_joint_reset_distribution}
\end{align}
The binomial factor counts the possible orders of the $m+n$ resetting cycles before the final absorbing cycle. Normalization follows from $\mathcal{G}(1,1)=1$, using $P_\mathrm{abs}+P_T+P_p=1$. 
The mean counts follow directly from the first derivatives of Eq.~\eqref{count_generating_function_expansion}
\begin{align}
    \left\langle\mathcal N_{th}\right\rangle
    &=\left.\frac{\partial\mathcal G}{\partial z_{th}}
    \right|_{(1,1)}
    =\frac{P_T}{P_\mathrm{abs}}=\frac{\widetilde j_\alpha(\lambda)}
    {\widetilde j_0(\lambda)},
    \label{mean_threshold_count_final}\\
    \left\langle\mathcal N_p\right\rangle
    &=\left.\frac{\partial\mathcal G}{\partial z_p}
    \right|_{(1,1)}
    =\frac{P_p}{P_\mathrm{abs}}=\frac{\lambda\widetilde q_0(\lambda)}
    {\widetilde j_0(\lambda)}
   =\lambda\left\langle T_\mathrm{tot}(u,\lambda)\right\rangle.
    \label{mean_poisson_count_final}
\end{align}
Thus, the mean number of resets of either type is the ratio of its single-cycle reset probability to the probability of target absorption.
In the limit of threshold resetting alone, we recover \cite{biswas2026optimal,Biswas2025_Threshold}
\begin{align}
    \lim_{\lambda\to0^+}
    \left\langle\mathcal N_{th}\right\rangle
    =\frac{\widetilde j_\alpha(0)}{\widetilde j_0(0)}
    =\frac{\epsilon_\alpha}{\epsilon_0}
    =\frac{u}{1-u},
    \label{mean_threshold_count_zero_rate}
\end{align}
which does not explicitly depend on the initial condition, the apex angle and more generally, the nature of the first-passage time density (also see \cite{pal2026resetting} for such nature of splitting probabilities). 

The asymptotic behavior the mean threshold and reset counts for large  $\lambda$ 
can be obtained by using the asymptotic values of boundary-currents  Eqs. (\ref{aymptotic_threshold_flux_zero}) and (\ref{aymptotic_threshold_flux_alpha}), derived using a saddle point integration in
Appendix~\ref{app:large-lambda-count}. Considering the expression
$\langle\mathcal N_{th}\rangle$ from
Eq.~\eqref{mean_threshold_count_final} and using  it \eref{aymptotic_threshold_flux_zero} and \eref{aymptotic_threshold_flux_alpha}, one has  
\begin{align}
   \langle\mathcal N_{th}\rangle
    &\sim
    \exp\Bigg[\sqrt{\frac{\lambda}{D}}2r_0\cos\left(\frac{\alpha}{2}\right)
    \sin\left[\alpha\left(u-\frac12\right)\right]\Bigg],~~\lambda>>0.
    \label{large_lambda_mean_th_counts_intext}
\end{align}

Further, considering Eqs.(\ref{aymptotic_threshold_flux_zero}) and (\ref{aymptotic_threshold_flux_alpha}),  and noticing that $\widetilde j_0(\lambda)$, and  $\widetilde j_\alpha(\lambda)$ approaches zero in the limit $\lambda\to\infty$,
$P_p=1-[\widetilde j_0(\lambda)+\widetilde j_\alpha(\lambda)]\to1$,
and Eq.~\eqref{mean_poisson_count_final} leads to the asymptotic expression$\langle\mathcal N_p\rangle\sim1/\widetilde j_0(\lambda)$ resulting in,
\begin{align}
    \langle\mathcal N_p\rangle
    &\sim
    \exp\left[\sqrt{\frac{\lambda}{D}}r_0\sin\theta_0\right],~~~\lambda>>0
    \label{large_lambda_mean_pr_counts_intext}
\end{align}
to the leading order, implying an exponential divergence in $\langle\mathcal N_p\rangle$  when $\lambda\to\infty$. The asymptotic behavior of the mean counts in Eqs. (\ref{large_lambda_mean_th_counts_intext}) and (\ref{large_lambda_mean_pr_counts_intext}) show that the mean threshold count $\langle\mathcal N_{th}\rangle$ tends to zero, equals one or diverges respectively for $u<1/2,~u=1/2$ and $u>1/2$, as governed by the sign of $\sin[\alpha(u-1/2)]$ whereas the mean Poissonian count diverges for any initial starting position inside the wedge as $r_0\sin\theta_0>0$.

\subsection{Reset-count correlation}
\label{reset_correlation}
Increasing the Poissonian resetting rate $\lambda$ can affect the probability of reaching the threshold as well the target absorption probability within a cycle, and therefore the number of cycles required on average before completion. 
To quantify how the two counts vary together across repeated cycles, we use Pearson's correlation coefficient, which measures the covariance of their deviations from their respective mean, normalized by their standard deviations. 

The Pearson's correlation coefficient for the reset-counts is be defined as follows,
\begin{align}
  \operatorname{Corr}(\mathcal N_{th},\mathcal N_p)
=
\frac{
\operatorname{Cov}(\mathcal N_{th},\mathcal N_p)
}{
\sqrt{
\operatorname{Var}(\mathcal N_{th})
\operatorname{Var}(\mathcal N_p)
}
}
\label{correlation_threshold_poisson}
\end{align}
where $\operatorname{Cov}(.)$ and $\operatorname{Var}(.)$ are the covariance and variance of the respectively. These follow from the first and second derivatives of the joint generating function in Eq.~\eqref{count_generating_function}, evaluated at $z_{th}=z_p=1$:
\begin{align}\label{forms_variance_covariance}
\operatorname{Var}(\mathcal N_{th})
&=
\frac{P_T(P_\mathrm{abs}+P_T)}{P_\mathrm{abs}^2},
&
\nonumber\\\operatorname{Var}(\mathcal N_p)
&=
\frac{P_p(P_\mathrm{abs}+P_p)}{P_\mathrm{abs}^2},~~~~~\rm{and}
\\
\operatorname{Cov}(\mathcal N_{th},\mathcal N_p)
&=
\frac{P_TP_p}{P_\mathrm{abs}^2}>0.\nonumber
\end{align}
By substituting \eref{forms_variance_covariance} in \eref{correlation_threshold_poisson} we obtain,
\begin{align}
    \operatorname{Corr}(\mathcal N_{th},\mathcal N_p)
    &=
    \sqrt{
        \frac{P_TP_p}
        {(P_\mathrm{abs}+P_T)(P_\mathrm{abs}+P_p)}
    }.
    \label{correlation_probability_form}
\end{align}
Using the mean-count expressions
$\langle\mathcal N_{th}\rangle=P_T/P_\mathrm{abs}$ and
$\langle\mathcal N_p\rangle=P_p/P_\mathrm{abs}$ from
Eqs.~\eqref{mean_threshold_count_final} and
\eqref{mean_poisson_count_final},  in  Eq.~\eqref{correlation_probability_form}, 
$\operatorname{Corr}(\mathcal N_{th},\mathcal N_p)$ can also be identified as
\begin{align}
\operatorname{Corr}(\mathcal N_{th},\mathcal N_p)
&=
\left(1+\frac{1}{\langle\mathcal N_{th}\rangle}\right)^{-1/2}
\left(1+\frac{1}{\langle\mathcal N_p\rangle}\right)^{-1/2}.
\label{correlation_mean_counts_exact}
\end{align}

\begin{figure}[htbp]
    \centering
    \includegraphics[width=\columnwidth]
    {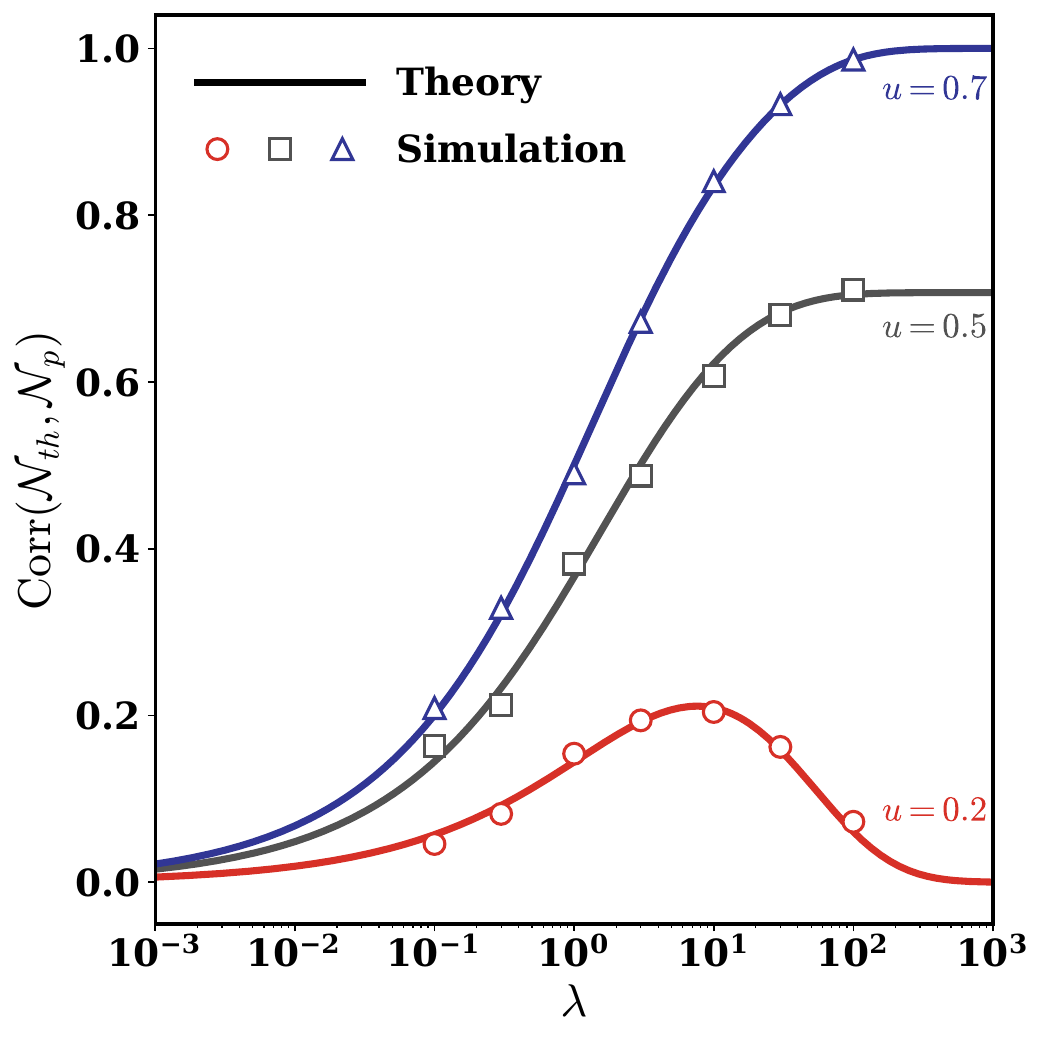}
    \caption{Pearson correlation between the numbers of
    threshold and Poissonian resets accumulated before target
    absorption, plotted against the Poissonian resetting rate
    $\lambda$ on a logarithmic scale. The analytical prediction
    (Eq.~\eqref{correlation_mean_counts_exact}) is plotted in solid lines 
    and the symbols represent the quantity calculated from
    direct simulation of independent
    Brownian searches  for
    $u=0.2$, $0.4$, $0.5$, and $0.7$,
    $r_0=D=1$, and $\alpha=\pi/3$, as a function of $\lambda$.}
    \label{fig:correlation_vs_lambda}
\end{figure}

To study the behavior of the correlation with resetting rate $\lambda$, using Eqs.~\eqref{mean_threshold_count_final} and \eqref{mean_poisson_count_final}, we express both mean reset counts $\langle\mathcal{N}_{th}\rangle$ and $\langle\mathcal{N}_{p}\rangle$ in 
Eq.~\eqref{correlation_mean_counts_exact}, in terms of the Laplace transform boundary currents $\widetilde j_0(\lambda)$ and $\widetilde j_\alpha(\lambda)$ [Eq.~\eqref{saddle_current}]. Behavior of the correlation against reset rate is  displayed in \fref{fig:correlation_vs_lambda}. In the same figure,  we also plot $\operatorname{Corr}(\mathcal N_{th},\mathcal N_p)$ vs $\lambda$,  for different initial positions $u$,  which illustrate distinct characteristics in their behavior.
For initial configurations closer to  the target ($u=0.2$ and $0.4$), the correlation initially increases, reaches a maximum, and then approaches zero. At the bisector of the wedge ($u=1/2$), it increases towards $(1/\sqrt{2})$, whereas for  $u$ beyond $1/2$, ($u=0.7$), the correlation coefficient tends towards unity. Thus in short, the number of both reset mechanisms are positively correlated but the response of increasing the Poissonian reset rate depends on the initial position.

The asymptotic behavior of
$\operatorname{Corr}(\mathcal N_{th},\mathcal N_p)$ for large
$\lambda$ follows those of the mean reset counts $\langle\mathcal N_{th}\rangle$ and $\langle\mathcal N_{p}\rangle$. As noted in
Eq.~\eqref{large_lambda_mean_pr_counts_intext},
$\langle\mathcal N_p\rangle\to\infty$ as $\lambda\to\infty$.
Substituting this large-rate behavior into
Eq.~\eqref{correlation_mean_counts_exact}, the correlation
takes the asymptotic form
\begin{align}
    \operatorname{Corr}(\mathcal N_{th},\mathcal N_p)
    &\sim
    \left(1+\frac{1}{\langle\mathcal N_{th}\rangle}\right)^{-1/2},
    \qquad \lambda>>0.
    \label{large_lambda_correlation_threset_intext}
\end{align}
to the leading order. For $u<1/2$, the mean threshold count
$\langle\mathcal N_{th}\rangle$ tends to zero as
$\lambda\to\infty$. In this regime, the factor on the
right-hand side of Eq.~\eqref{large_lambda_correlation_threset_intext}
can be written as
$\sqrt{\langle\mathcal N_{th}\rangle}
(1+\langle\mathcal N_{th}\rangle)^{-1/2}$.
Expanding the second factor for small
$\langle\mathcal N_{th}\rangle$ gives the leading
large-$\lambda$ behavior,
\begin{align}
    \operatorname{Corr}(\mathcal N_{th},\mathcal N_p)
    &\sim \sqrt{\langle\mathcal N_{th}\rangle},
    \qquad 0<u<\frac12.
    \label{large_lambda_correlation_small_count}
\end{align}
At $u=1/2$, the mean threshold count tends to unity, $\langle \mathcal{N}_{th}\rangle=1$, with the correlation approaching $1/\sqrt{2}$ as $\lambda\to\infty$. Further, for $u>1/2$, the mean threshold count diverges. Expanding the leading factor in powers of $1/\langle \mathcal{N}_{th}\rangle$ results in the following asymptotic form for the correlation 
\begin{align}
    \operatorname{Corr}(\mathcal N_{th},\mathcal N_p)
    &\sim 1-\frac{1}{2\langle\mathcal N_{th}\rangle},
    \qquad \frac12<u<1.
    \label{large_lambda_correlation_large_count}
\end{align}

Summarizing the three limiting behaviors obtained above, 
\begin{align}
    \lim_{\lambda\to\infty}
    \operatorname{Corr}(\mathcal N_{th},\mathcal N_p)
    &=
    \begin{cases}
        0, & 0<u<1/2,\\
        1/\sqrt{2}, & u=1/2,\\
        1, & 1/2<u<1,
    \end{cases}
\end{align}
where the three regimes correspond to reset location near to
the target wall, on the wedge bisector, or the
threshold boundary, respectively.

\section{Conclusion}\label{sec:conclusion}
We study the first passage properties of a Brownian target search in two-dimensional wedge under threshold resetting, both without and with Poissonian resetting. The threshold resetting mechanism realizes a scenario where the effect confinement geometry can influence statistical properties such as search time to targets. We arrive at an exact expression for the mean first passage time towards the target boundary. Under threshold resetting, for a fixed initial radius, the mean first passage time increases monotonically with the searchers distance $u$, from the target boundary. For wedge angles $\alpha<\pi/2$, it vanishes linearly near the target and approaches a finite limiting value as it approaches the threshold boundary. This finite limiting value has a simple physical interpretation: Although the mean number of threshold resets increases as the reset point approaches the threshold boundary, the mean time to reach either boundary and the probability of reaching the target within a single cycle both vanish in proportion to $(1-u)$. The near-target MFPT analysis reveals that when the searcher is very close to the target, the system resembles a one-dimensional interval search, with an effective size of the equivalent one-dimensional interval strictly dictated by the size of the wedge. For $\alpha\to\pi/2$, the effective size of the corresponding $1D$ interval approaches infinity, so does the MFPT. 

We also derive an exact expression for the second moment, which is finite for $0<\alpha<\pi/4$.
For $\pi/4\leq\alpha<\pi/2$, the search has a finite mean but divergent variance, showing that rare, long searches remain important despite a finite average completion time. The long-time analysis of the first-passage time distribution under TR, exhibits an algebraic tail $f_{TR}(t) \sim t^{-(1+\pi/2\alpha)}$ equivalent to the non-reset wedge domain. Thus the large-time algebraic tail, is preserved under the threshold resetting renewal structure. The physical interpretation for this is as follows: the long-time deep radial excursions, accumulate more time without hitting either boundaries, thus reducing the contributions to the MFPT due to threshold resets, making it equivalent to the non-reset wedge with absorbing boundaries. However, threshold resetting changes the weight of the long searches through the reciprocal factor $1-u$ (\eref{long_tails_of_Sandf}), without removing the algebraic decay associated with the unbounded radial direction.

In addition to the boundary induced threshold resetting, the Poissonian resets to the initial point also was introduced.
The addition of resetting within the domain, interrupts the statistics of the boundary hitting, thus, both target search time and the threshold resets events experience interruptions. To study the combined effect of these two reset events on the search time statistics, MFPT under the combined reset protocol was derived. To assess whether an additional stochastic (Poissonian) resetting optimizes these search times, we evaluated the coefficient of variation ($CV_{TR}$) with the diffusion under threshold resetting taken as the underlying process.  The study confirms a behavioral transition, with the stochastic resetting lowering the mean search time in specific parametric region which is quantified.

Our study shows that the TR and PR events are non-trivially correlated. To arrive there, we employ a renewal approach to compute the joint count statistics of number of reset intervention, prior to the completion of the first-passage process. Finally, to estimate the correlation, we estimate the Pearson's correlation coefficient for the reset counts. We find that the two reset counts have a positive Pearson correlation, with all parameters kept identical. Quite interestingly, at large Poisson rate, we observe an interesting crossover in the Pearson correlation coefficient which depends on the initial/resetting position quite non-trivially. In particular, we show that for wedges with $\alpha=\pi/n<\pi/2$, the correlation decays to zero for initial position in the lower half of the wedge $u<1/2$. If the initial position is on the wedge bisector $u=1/2$, the correlation converges to $1/\sqrt{2}$ and saturates to unity for $u>1/2$. Since the cost  or resource functions depend on the count statistics, one would anticipate such crossover instead of a monotonically increasing correlation as Poisson rate is increased. That such measured probing could forestall a larger loss is an interesting direction that we would like to pursue in future.
 
Possible extensions include looking into wedge model with radial drift from its apex~\cite{krapivsky2022}. Another realistic aspect would be to fix finite return times instead of the instantaneous jumps. This return to the initial state can be optimized with different modes of reset transport, which depend on the landscape of the domain. Coupling either reset mechanism to an
internal coordinate may connect the framework to neuronal firing, cellular
regulation, and reaction-coordinate dynamics
\cite{kracikova2013threshold,burkitt2006review,ray2021resetting}. Resetting can involve thermodynamic cost
\cite{fuchs2016stochastic,gupta2020workfluctuations,pal2021thermodynamic,entropy_resetting},
and assigning a cost to each reset allows these interventions
to be included in the thermo-kinetic optimization of search strategies. This remains an interesting question for future studies. Concluding, we believe that our framework will be useful for systematically investigating the interplay of competing resetting mechanisms in first-passage processes across more general geometries and higher dimensions.

\begin{acknowledgments}
AP acknowledges research funding under the scheme
ANRF/ARGM/2025/001623 from ANRF, India and research support from the Department of Science and Technology, India, SERB Start-up Research Grant Number SRG/2022/000080. AP also acknowledges the International Research Project (IRP) titled ``Classical and quantum dynamics in out of equilibrium systems'' by CNRS, France. Finally, AP acknowledges research support from the Department of Atomic Energy via the SoftMatter Apex projects. 
\end{acknowledgments}
\appendix

\section{Renewal formalism for threshold-resetting moments}
\label{app:renewal_moments}
To obtain expressions for the first and  the second moments 
of the diffusing particle under the threshold reset target search model, we start with
the renewal relation under threshold resetting,

\begin{align}\label{laplace_renewal}
\widetilde f_{TR}(s)
=
\frac{\widetilde j_0(s)}
{1-\widetilde j_\alpha(s)},
\end{align}
where the quantities on the LHS involve the Laplace transform~[see \eref{laplace_transform_convention}] of the first-passage density under threshold resetting~[Eq.~\eqref{threshold_survival_density_definition}] and RHS the Laplace transforms~ of currents in the absence of resetting~[\eref{boundary_current_definitions}]. 
Expanding first passage time density and and the boundary fluxes near $s=0$, 
\begin{align}\label{threshold_fpt_expansion}
\widetilde f_{TR}(s)
&=
1
-s\left\langle T_{TR}\right\rangle
+\frac{s^2}{2}\left\langle T_{TR}^2\right\rangle
+o(s^2),
\\
\widetilde j_i(s)
&=
\epsilon_i
-sm_{1,i}
+\frac{s^2}{2}m_{2,i}
+o(s^2).
\label{current_moment_expansion}
\end{align}
where we have used $m_{n,i}=\int_{0}^{\infty}dt~t^n J_i(t)$ [also defined in the main text  at Eq.~\eqref{partial_moment_definition}], and that $\widetilde j_i(0)=\epsilon_i,~\{i=0,\alpha\}$, the splitting probabilities~[Eq.~\eqref{model_splitting_probabilities}] . For the unbounded wedge, the first moment is finite for
$\alpha<\pi/2$, while the second moment is finite for
$\alpha<\pi/4$. The second-order expansion below is therefore
finite and valid in the domain
\begin{align}
0<\alpha<\frac{\pi}{4}.
\end{align}

Substituting Eq.~\eqref{current_moment_expansion} into
Eq.~\eqref{laplace_renewal}, and using
$1-\epsilon_\alpha=\epsilon_0$, gives
\begin{align}\label{renewal_moment_ratio}
1
-s\left\langle T_{TR}\right\rangle&
+\frac{s^2}{2}\left\langle T_{TR}^2\right\rangle
+o(s^2)
\nonumber\\=&
\frac{
\epsilon_0
-sm_{1,0}
+\dfrac{s^2}{2}m_{2,0}
+o(s^2)
}{
\epsilon_0
+sm_{1,\alpha}
-\dfrac{s^2}{2}m_{2,\alpha}
+o(s^2)
}.
\end{align}
Multiplying both sides of Eq.~\eqref{renewal_moment_ratio} by its
denominator on R.H.S and equating the coefficients of order $s$ we have, 
\begin{align}
\epsilon_0\left\langle T_{TR}\right\rangle
=
m_{1,0}+m_{1,\alpha}.
\label{app-a-moment-sum1}
\end{align}
Further, from Eq.~\eqref{partial_moment_sum} with substitution $n=1$, one has
\begin{align}
m_{1,0}+m_{1,\alpha}
=
\left\langle T_0\right\rangle,
\label{app-a-moment-sum2}
\end{align}
and a direct comparison between \eref{app-a-moment-sum1} and \eref{app-a-moment-sum1} results in
\begin{align}\label{app_mean_t}
\left\langle T_{TR}\right\rangle
=
\frac{\left\langle T_0\right\rangle}{\epsilon_0}.
\end{align}

In the same way, equating the coefficients of order $s^2$ in \eref{renewal_moment_ratio} results in the relation for the second moment, 
\begin{align}
\epsilon_0
\left\langle T_{TR}^2\right\rangle
=
m_{2,0}
+
m_{2,\alpha}
+
2m_{1,\alpha}
\left\langle T_{TR}\right\rangle.
\label{app-a-second-moment-equation}
\end{align}
Further, substituting  \eref{partial_moment_sum} for n=2  and  Eq.~\eqref{app_mean_t} in \eref{app-a-second-moment-equation}, we obtain
\begin{align}\label{app_second_moment_corrected}
\left\langle T_{TR}^2\right\rangle
=
\frac{\left\langle T_0^2\right\rangle}{\epsilon_0}
+
\frac{
2m_{1,\alpha}\left\langle T_0\right\rangle
}{
\epsilon_0^2
}.
\end{align}

\subsection{Determining the boundary-resolved first moments}

To evaluate $m_{1,\alpha}$, we restore the dependence of the threshold-current transform, writing, $\widetilde j_\alpha(s;r_0,\theta_0)$. It satisfies the backward equation\cite{Krapivsky_2018,REDNER2023}
\begin{align}
D\nabla^2\widetilde j_\alpha
=
s\widetilde j_\alpha,
\end{align}
with boundary conditions
\begin{align}
\widetilde j_\alpha(s;r_0,0)
&=
0,
\qquad
\widetilde j_\alpha(s;r_0,\alpha)
=
1.
\end{align}
Its small-$s$ expansion is
\begin{align}
\widetilde j_\alpha(s;r_0,\theta_0)
=
\epsilon_\alpha(\theta_0)
-
s\,m_{1,\alpha}(r_0,\theta_0)
+
o(s).
\end{align}
At order $s^0$, the backward equation recovers the splitting probabilities $\epsilon_\alpha=u$ and $\epsilon_0=1-u$ already obtained in Eq.~\eqref{split_u}.

At order $s^1$, the backward equation gives
\begin{align}\label{partial_first_moment_pde}
D\nabla^2m_{1,\alpha}(r_0,\theta_0)
=
-\frac{\theta_0}{\alpha},
\end{align}
with
\begin{align}
m_{1,\alpha}(r_0,0)
=
m_{1,\alpha}(r_0,\alpha)
=
0.
\end{align}

Due to radial scale invariance of the wedge, the form of the solution is~\cite{redner2001guide,majumder2015}
\begin{align}
m_{1,\alpha}(r_0,\theta_0)
=
\frac{r_0^2}{4D}\,f(\theta_0).
\label{scaling_separable_form_partial_moment_1}
\end{align}
Substitution into Eq.~\eqref{partial_first_moment_pde} gives
\begin{align}
f''(\theta_0)
+
4f(\theta_0)
=
-\frac{4\theta_0}{\alpha},
\end{align}
with
\begin{align}
f(0)=f(\alpha)=0.
\end{align}
The solution is
\begin{align}
f(\theta_0)
=
\frac{\sin(2\theta_0)}{\sin(2\alpha)}
-
\frac{\theta_0}{\alpha}.
\end{align}
Therefore,
\begin{align}\label{m1alpha_exact}
m_{1,\alpha}(r_0,\theta_0)
=
\frac{r_0^2}{4D}
\left[
\frac{\sin(2\theta_0)}{\sin(2\alpha)}
-
\frac{\theta_0}{\alpha}
\right].
\end{align}

By reflection about the wedge bisector $\theta=\alpha/2$, which maps
$u\mapsto1-u$ \cite{majumder2015}. Therefore, $m_{1,\alpha}(u)=m_{1,0}(1-u).$
Hence we obtain,
\begin{align}\label{m10_exact}
m_{1,0}(r_0,\theta_0)
=
\frac{r_0^2}{4D}
\left[
\frac{\sin\!\left(2\alpha-2\theta_0\right)}
{\sin(2\alpha)}
-
\left(
1-\frac{\theta_0}{\alpha}
\right)
\right].
\end{align}
Substituting the expressions of $\langle T_0\rangle$,$\langle T_0^2\rangle$ from \cite{3bc4-nlkt}, $\epsilon_0$ and $m_{1,\alpha}$ in Eq.~\eqref{app_second_moment_corrected} we obtain the exact expression of the second moment of the first passage time under threshold reset,
\begin{align}
\left\langle T_{TR}^2\right\rangle
&=\frac{r_0^4}{96D^2(1-u)}
\Biggl[\frac{\cos\!\left(2\alpha(1-2u)\right)}{\cos(2\alpha)}
\nonumber\\
&\qquad-4\frac{\cos\!\left(\alpha(1-2u)\right)}{\cos\alpha}
+3\Biggr]
\nonumber\\
&\quad+\frac{r_0^4}{8D^2(1-u)^2}
\left[\frac{\cos\!\left(\alpha(1-2u)\right)}{\cos\alpha}-1\right]
\nonumber\\
&\qquad\times\left[\frac{\sin(2\alpha u)}{\sin(2\alpha)}-u\right].
\label{threshold_second_moment_explicit_corrected}
\end{align}

\subsection{Coefficient of variation under threshold resetting}
\label{app:threshold_cv}

Using $B_1,B_2$ and $B_3$ defined in Eq.~\eqref{B_functions_definition} and the additional dimensionless function,
\begin{align}\label{B4_definition}
B_4(u,\alpha)
&=
\frac{\sin\!\left(2\alpha(1-u)\right)}
{\sin(2\alpha)}
-(1-u)
\end{align}
we can compactly represent the first two non-reset moments and the two boundary-resolved
first moments as
\begin{align}
\left\langle T_0\right\rangle
&=
\frac{r_0^2}{4D}B_1(u,\alpha),
\\
\left\langle T_0^2\right\rangle
&=
\frac{r_0^4}{96D^2}B_2(u,\alpha),
\\
m_{1,\alpha}
&=
\frac{r_0^2}{4D}B_3(u,\alpha),
\\
m_{1,0}
&=
\frac{r_0^2}{4D}B_4(u,\alpha),
\end{align}
respectively.

Using these notation in Eqs.~\eqref{exact_T_TR} and
\eqref{threshold_second_moment_explicit_corrected}, and using the definition of the coefficient of variation in Eq.~\eqref{cv_definition}, gives
\begin{align}\label{threshold_cv_compact}
\operatorname{CV}_{TR}^2
&=
\frac{1-u}{6}
\frac{B_2(u,\alpha)}
{B_1(u,\alpha)^2}
+
2\frac{B_3(u,\alpha)}
{B_1(u,\alpha)}
-1.
\end{align}
Equation~\eqref{threshold_cv_compact} is finite for $0<\alpha<\frac{\pi}{4}.$
For $\pi/4\leq\alpha<\pi/2$, the mean remains finite but the second
moment diverges, and hence the coefficient of variation as well.

\section{Asymptotics of the first-passage-time density under threshold resetting}
\label{app:laplace_branch_cut}
To determine the long-time behavior of $S_{TR}(t)$
through the renewal relation in
Eq.~\eqref{renewal_laplace_intext}, we examine the
small-$s$ behavior of the non-reset quantities
$\widetilde q_0(s)$ and $\widetilde j_\alpha(s)$.
Their leading nonanalytic contributions follow from
the non-reset survival tail in
Eq.~\eqref{long_time_survival} and the boundary-current
asymptotics ~\cite{majumder2015},
\begin{align}
    J_0(t)\sim J_\alpha(t)
    \sim\frac{\beta A_0}{2}t^{-\beta-1}.
    \label{current_long_time}
\end{align}

Using the the Laplace-transform convention in  Eq.~\eqref{laplace_transform_convention} for $\widetilde q(s)$, we split the time integration limits at $t_D$, as defined in
Sec.~\ref{tail_analysis}, into the interval $(0,t_D]$ and $[t_D,\infty)$. Over the latter interval, we integrate using the leading long-time form of $Q_0(t)$ from Eq.~\eqref{long_time_survival}. Separating the regular powers of $s$ from the non-analytic contribution, we obtain
\begin{align}
    \widetilde q_0(s)
    &=
    \bigl(\widetilde q_0(s)\bigr)_{\mathrm{an}}
    \nonumber\\
    &\quad+
    \begin{cases}
        Cs^{\beta-1}+o(s^{\beta-1}),
        & \beta\notin\mathbb N,
        \\[2mm]
        \begin{aligned}
            &C's^{\beta-1}\log(st_D)
            \\
            &\quad+
            o\!\left(s^{\beta-1}|\log(st_D)|\right),
        \end{aligned}
        & \beta\in\mathbb N
    \end{cases}.
    \label{small_s_survival_laplace}
\end{align}
Similarly, for the threshold-current one obtains 
\begin{align}
    \widetilde j_\alpha(s)
    &=
    \bigl(\widetilde j_\alpha(s)\bigr)_{\mathrm{an}}
    \nonumber\\
    &\quad+
    \begin{cases}
        Bs^\beta+o(s^\beta),
        & \beta\notin\mathbb N,
        \\[2mm]
        \begin{aligned}
            &B's^\beta\log(st_D)
            \\
            &\quad+
            o\!\left(s^\beta|\log(st_D)|\right),
        \end{aligned}
        & \beta\in\mathbb N
    \end{cases}.
    \label{small_s_flux}
\end{align}
Here $\bigl(\widetilde q_0(s)\bigr)_{\mathrm{an}}$ and
$\bigl(\widetilde j_\alpha(s)\bigr)_{\mathrm{an}}$ collect the
analytic contributions in the respective small-$s$ expansions,
namely the regular integer powers of $s$ and $\mathbb N$ denote the set of all natural numbers. The coefficients of the nonanalytic terms are
\begin{align}
    C&=A_0\Gamma(1-\beta),
    \qquad
    B=-\frac{C}{2},
    \qquad \beta\notin\mathbb N,
    \nonumber\\
    C'&=\frac{(-1)^mA_0}{(m-1)!},
    \qquad
    B'=-\frac{C'}{2},
    \qquad \beta=m\in\mathbb N.
    \label{tail_transform_coefficients}
\end{align}
The remainders include the higher-order contributions from the
non-reset tails.
Substituting the preceding expansions into the
renewal relation [Eq.~\eqref{renewal_laplace_intext}], therefore gives
\begin{align}
    \widetilde S_{TR}(s)
    &=
    \frac{
        \bigl(\widetilde q_0(s)\bigr)_{\mathrm{an}}
    }{
        1-\bigl(\widetilde j_\alpha(s)\bigr)_{\mathrm{an}}
    }
    \nonumber\\
    &\quad+
    \begin{cases}
        \displaystyle
        \frac{C}{1-u}s^{\beta-1}
        +o(s^{\beta-1}),
        & \beta\notin\mathbb N,
        \\[3mm]
        \begin{aligned}
            &\frac{C'}{1-u}s^{\beta-1}\log(st_D)
            \\
            &\quad+
            o\!\left(s^{\beta-1}|\log(st_D)|\right),
        \end{aligned}
        & \beta\in\mathbb N
    \end{cases}.
    \label{tail_reset_survival_small_s}
\end{align}
The renewal structure therefore multiplies the leading
nonanalytic coefficient by $1/(1-u)$, without changing its power.

Since $\widetilde f_{TR}(s)=
    1-s\widetilde S_{TR}(s)$, the threshold first passage density can be expanded as, 
\begin{align}
    \widetilde f_{TR}(s)
    &=
    1+\cdots
    \nonumber\\
    &\quad+
    \begin{cases}
        \mathcal A_1s^\beta+o(s^\beta),
        & \beta\notin\mathbb N,
        \\[2mm]
        \begin{aligned}
            &\mathcal A_2s^\beta\log(st_D)
            \\
            &\quad+
            o\!\left(s^\beta|\log(st_D)|\right),
        \end{aligned}
        & \beta\in\mathbb N
    \end{cases},
    \label{app_threshold_density_small_s}
\end{align}
where the dots denote the regular powers associated with the
finite moments, and
\begin{align}
    \mathcal A_1
    &=-\frac{C}{1-u},
    \qquad \beta\notin\mathbb N,
    \label{app_density_coefficient_noninteger}\\
    \mathcal A_2
    &=-\frac{C'}{1-u},
    \qquad \beta\in\mathbb N.
    \label{app_density_coefficient_integer}
\end{align}
These are the coefficients of the leading nonanalytic
terms and contain the geometric and dimensional factors.

The connection between the small-$s$ singularities and the
long-time tail is described by Abelian--Tauberian theory
\cite{Feller1971}. For the density, asymptotic Laplace inversion
of the fractional-power and logarithmic branch terms gives the
same algebraic decay in both cases
\cite{Doetsch1974},
\begin{align}
    f_{TR}(t)
    &\sim
    \mathcal C\,t^{-\beta-1},
    \qquad t\to\infty,
    \label{app_threshold_density_tail}\\
    \mathcal C
    &=
    \frac{\beta A_0}{1-u}>0.
    \label{app_threshold_tail_amplitude}
\end{align}
Equivalently,
\begin{align}
    f_{TR}(t)
    &\sim
    \mathcal C\,
    t^{-\left(1+\frac{\pi}{2\alpha}\right)}.
    \label{compact_threshold_tail_alpha}
\end{align}
Integrating this density from the observation time to infinity
gives
\begin{align}
    S_{TR}(t)
    &=
    \int_t^\infty f_{TR}(v)\,dv
    \sim
    \frac{A_0}{1-u}\,t^{-\beta}.
    \label{tail_threshold_survival}
\end{align}
Threshold resetting therefore changes the tail amplitude while
preserving the algebraic exponent set by the wedge angle.

\section{Large lambda asymptotics of average reset counts and their correlation}
\label{app:large-lambda-count}
We derive the reset rate $\lambda\to\infty$ behavior of the average reset counts and their correlation  from the finite-image probability current expressions. We take the limit \(\lambda\to\infty\) at a fixed reset point and restrict the calculation to acute finite-image wedges, \(\alpha=\pi/n\) with \(n\ge3\). We use $\kappa=\sqrt{\lambda/D}$ throughout. 

The Laplace transforms of the boundary current is,
\begin{align}
   \widetilde j_\theta(s)
    &=
    \sigma_\theta\frac{r_0\kappa}{\pi}
    \sum_{k=0}^{n-1}
    \sin\phi_{\theta,k}
    \int_0^\infty
    \frac{K_1\!\left(\kappa\rho_{\theta,k}(r)\right)}
    {\rho_{\theta,k}(r)}\,dr.
    \label{saddle_current}
\end{align}
where we have used the following condensed notations, 
\begin{align}
\sigma_\theta&=
\begin{cases}
+1,&\theta=0,\\
-1,&\theta=\alpha,
\end{cases}\nonumber\\
    \phi_{\theta,k}
    &=2k\alpha+\theta_0-\theta,
    \nonumber\\
    \rho_{\theta,k}(r)
    &=\left[r^2+r_0^2-2rr_0\cos\phi_{\theta,k}\right]^{1/2}.
    \label{saddle_image_definitions}
\end{align}
Here $r$ varies along the target ray, while the image indexed by $k$
remains fixed. 

To demonstrate the calculation, we focus on the target boundary current first at $\theta=0$.
The modified bessel function of the second kind $K_1(z)$ has the following large-argument expansion,
\begin{align}
    K_1(z)=\sqrt{\frac{\pi}{2z}}e^{-z}[1+O(z^{-1})],\qquad z\to \infty.
\end{align}
Therefore, the radial integrand in the current expression Eq.~\eqref{saddle_current} at $\theta=0$,
for large-$\kappa$ becomes,
\begin{align}
    \frac{K_1(\kappa\rho_{0,k}(r))}{\rho_{0,k}(r)}
    &\sim
    \sqrt{\frac{\pi}{2\kappa}}\,
    \rho_{0,k}(r)^{-3/2}
    e^{-\kappa\rho_{0,k}(r)}.
    \label{large_lambda_bessel}
\end{align}
The resulting radial integral is,
\begin{align}
    I_k(\kappa)=\int_0^\infty\rho_{0,k}(r)^{-3/2}e^{-\kappa\rho_{0,k}(r)}~dr.
    \label{radial_integral_part}
\end{align}
As $\kappa\to \infty$, the exponential function localizes the integral $I_k(\kappa)$ near the minimum of $\rho_{0,k}(r)$. Here $\rho_{0,k}(r)^{-3/2}$ is $\kappa$-independent term that is finite within the integration limit. From the form of $\rho_{0,k}(r)$ it is evident that, for acute wedges($\alpha<\pi/2$), has no roots for $r\in(0,\infty)$ and is continuously differentiable. Consequently, to solve the integral \eqref{radial_integral_part}, we can apply the Laplace's method, which, for a unique global minimum exists at some stationary point $r^*>0\;\text{for}\;\rho_{0,k}(r)$ gives~\cite{Olver1974}
\begin{align}
    \int_0^\infty \rho_{0,k}(r)^{-3/2}&e^{-\kappa\rho_{0,k}(r)}\, dr
    \sim\nonumber \\
    &\rho_{0,k}(r^*)^{-3/2}e^{-\kappa\rho_{0,k}(r^*)}
    \sqrt{
        \frac{2\pi}{\kappa\rho_{0,k}''(r^*)}
    },
    \quad \kappa\to\infty.
    \label{laplace_method}
\end{align}
We therefore locate $r^*$ for which $\rho_{i,k}(r^*)$ is a minimum for the corresponding $k^{th}-$image term.
For the finite-image angles considered here,
$\cos\theta_0>\cos\phi_{0,k}$ for $k=1,\ldots,n-1$.
Consequently, the $k=0$ image has the smallest distance to the target
ray. Its unique interior minimum satisfies
\begin{align}
    r^*&=r_0\cos\theta_0,
    \nonumber\\
    \rho_{0,0}(r^*)&=r_0\sin\theta_0,
    \qquad
    \rho_{0,0}''(r^*)=\frac{1}{r_0\sin\theta_0}>0.
    \label{app_dominant_image_minimum}
\end{align}
All other images have strictly larger minimum distances, including
those whose closest point on the ray is the apex, at distance $r_0$.
Their contributions are therefore exponentially smaller at fixed
$0<\theta_0<\alpha$.

Substituting these values into the Laplace approximation
\eqref{laplace_method}\cite[Chap.~3, pp.~80--88]{Olver1974}
and restoring the Bessel and current prefactors gives
\begin{align}
    \widetilde j_0(\lambda)
    &\sim
    \exp\!\left[-r_0\sqrt{\frac{\lambda}{D}}\sin\theta_0\right],
    \qquad \lambda\to\infty.
    \label{aymptotic_threshold_flux_zero}
\end{align}
Reflection about the wedge bisector replaces $\theta_0$ by
$\alpha-\theta_0$, giving the threshold current
\begin{align}
    \widetilde j_\alpha(\lambda)
    &\sim
    \exp\!\left[-r_0\sqrt{\frac{\lambda}{D}}
    \sin(\alpha-\theta_0)\right],
    \qquad \lambda\to\infty.
    \label{aymptotic_threshold_flux_alpha}
\end{align}

Both currents vanish, while probability conservation within a cycle
gives $\lambda\widetilde q_0(\lambda)
=1-\widetilde j_0(\lambda)-\widetilde j_\alpha(\lambda)\to1$.
Substituting these asymptotic forms into
Eq.~\eqref{correlation_mean_counts_exact} yields
\begin{align}
    \operatorname{Corr}(\mathcal N_{th},\mathcal N_p)
    &\sim
    \Biggl\{1+\exp\Biggl[-2r_0\sqrt{\frac{\lambda}{D}}
    \nonumber\\
    &\quad\times\cos\!\left(\frac{\alpha}{2}\right)
    \sin\!\left[\alpha\left(u-\frac12\right)\right]
    \Biggr]\Biggr\}^{-1/2}.
    \label{large_lambda_correlation}
\end{align}
Since $\sin[\alpha(u-1/2)]$ has the sign of $u-1/2$, the
large-rate limits are
\begin{align}
    \lim_{\lambda\to\infty}
    \operatorname{Corr}(\mathcal N_{th},\mathcal N_p)
    =
    \begin{cases}
        0, & 0<u<1/2,\\[1mm]
        1/\sqrt{2}, & u=1/2,\\[1mm]
        1, & 1/2<u<1
    \end{cases}.
    \label{app_large_lambda_correlation_limits}
\end{align}
These limits explain the large-rate behavior of the curves in
Fig.~\ref{fig:correlation_vs_lambda}.

\section{Numerical simulations and data generation}\label{numerical_simulations}
We simulate independent searches in the wedge $0<\theta<\alpha$, starting
from the reset point $\boldsymbol r_0=(r_0\cos\theta_0,r_0\sin\theta_0)$,
where $u=\theta_0/\alpha$. A search includes all reset events and ends at
target absorption. The same diffusion and boundary-event routines(Reset or absorption) are
used for the mean first-passage times, survival probabilities, and reset counts.
\subsection{Brownian search dynamics and boundary detection}
We discretize the Brownian dynamics in Eq.~\eqref{trajectory_langevin} starting from $(x_0,y_0)=(r_0\cos\theta_0,r_0\sin\theta_0)$. The numerical update of each coordinate, at an arbitrary step $j$, is,
\begin{align}
    x_{j+1}=x_j+\sqrt{2D\Delta t} \xi^{(x)}_j \nonumber\\
    y_{j+1}=y_j+\sqrt{2D\Delta t} \xi^{(y)}_j
\end{align}
where $ \xi^{(x)}_j$ and $ \xi^{(y)}_j$ are independent standard normal variables. 
\subsubsection{Boundary detection}\label{boundary_detection}
 Inside the wedge, the signed normal distances to the target and threshold
lines are $d_{0}=y_j$ and $d_\alpha=x_j\sin\alpha-y_j\cos\alpha$, respectively. A non-positive distance triggers the corresponding boundary event. If $d_0\le0,d_\alpha>0$, triggers the absorption event the search ends and if $d_0>0,d_\alpha\le0$, then the threshold reset triggers and the cycle restarts. When both are positive $d_0,d_\alpha>0$, then the process is inside the wedge. Another case is when both $d_0,d_\alpha\le0$, since this can trigger both boundary events, we propose the following implementation; Suppose before triggering the boundary distances negative $d_0,d_\alpha\le0$, the coordinate was $(x_1,y_1)$ and $(x_2,y_2)$ after triggering that condition. We can calculate the slope of this path line segment for that particular step jump,
$$m_s=\frac{y_2-y_1}{x_2-x_1}.$$
The jump causes the segment to cross the horizontal absorbing boundary at some point $(x_{abs},0)$. The sign of $x_{abs}$ determines whether the absorbing boundary was triggered first or the threshold boundary.  This is implemented as follows
\begin{align}
    x_{abs}=\frac{m_sx_1-y_1}{m_s}.
\end{align}
If,
\begin{itemize}
        \item $x_{abs}<0$ threshold boundary crossed first
        \item $x_{abs}>0$ absorbing boundary crossed first
        \item $x_{abs}=0$ crossed the apex
    \end{itemize}
This determines the boundary crossing criterion in all simulations used.
\subsection{Threshold resetting}
A target-boundary event terminates the search. A threshold-boundary event
instantaneously returns the particle to $\vec r_0$ and increments
the threshold-reset count by one. The accumulated search time is retained,
and subsequent diffusion uses fresh random increments. This protocol alone
is used for the threshold-reset MFPT and survival calculations.
\subsection{Poissonian resetting}
For the combined protocol, we sample the exponential waiting time $R$ defined in Eq.~\eqref{combined_poisson_density} for each reset clock. Each particle receives its own sample waiting time, drawn from the same exponential distribution with mean $1/\lambda$. We draw a uniform random number $U\in(0,1)$ and generate the waiting time 
$R$ as,
\begin{align}
    R=-\frac{\log(1-U)}{\lambda}
    \label{exponential_waiting_time}
\end{align}
When the search begins, we draw $R$. The Brownian motion advances in steps of size $\Delta t$ subject to boundary events. After $j$ full steps since the clock was drawn, provided no earlier event has ended the cycle, the remaining time is,
\begin{align}
    t_{rem}=R-j\Delta t.
\end{align}
When $t_{rem}\geq0$ the simulation, at each step, checks the minimum of the remaining time and  updates the time step size as,
\begin{align}
    \Delta t\to\hat\Delta t=\min(\Delta t,t_{rem}).
\end{align}
If at $(j+1)^{th}-$step, $\hat\Delta t=\Delta t$, then the particle will continue undergoing diffusion, and if $\hat\Delta t=t_{rem}$. In case $\Delta t=t_{rem}$ still reset is triggered. The last Brownian step will increments the independent coordinates$(i\in(x,y))$ value by an amount $\sqrt{2D\hat\Delta t}\times\xi_i$  and reset will be triggered, updating the coordinates to the initial value $\vec r\to\vec r_0$. At each step, the boundary event is also monitored. To evaluate MFPT for the corresponding search, the total accumulated time is recorded, till the occurence of the boundary event corresponding to absorption.

\bibliography{bibthreshold_draft_v04}

@book{redner2001guide,
  author        = {Redner, Sidney},
  title         = {A Guide to First-Passage Processes},
  publisher     = {Cambridge University Press},
  address       = {Cambridge},
  year          = {2001},
  doi           = {10.1017/cbo9780511606014},
  isbn          = {9780521652483}
}

@book{metzler2014first,
  editor        = {Metzler, Ralf and Oshanin, Gleb and Redner, Sidney},
  title         = {First-Passage Phenomena and Their Applications},
  publisher     = {World Scientific},
  address       = {Singapore},
  year          = {2014},
  doi           = {10.1142/9104},
  isbn          = {9789814590280}
}

@article{ahmad2019first,
  title={First passage of a particle in a potential under stochastic resetting: A vanishing transition of optimal resetting rate},
  author={Ahmad, Saeed and Nayak, Indrani and Bansal, Ajay and Nandi, Amitabha and Das, Dibyendu},
  journal={Physical Review E},
  volume={99},
  number={2},
  pages={022130},
  year={2019},
  publisher={APS}
}

@article{pal2021thermodynamic,
  title={Thermodynamic uncertainty relation for systems with unidirectional transitions},
  author={Pal, Arnab and Reuveni, Shlomi and Rahav, Saar},
  journal={Physical Review Research},
  volume={3},
  number={1},
  pages={013273},
  year={2021},
  publisher={APS}
}

@article{pal2026resetting,
  title={Resetting optimized competitive first-passage outcomes in non-Markovian systems},
  author={Pal, Suvam and Das, Rahul and Pal, Arnab},
  journal={APS Open Science},
  volume={1},
  pages={000155},
  year={2026},
  publisher={APS}
}

@incollection{pal2024random,
  title={Random resetting in search problems},
  author={Pal, Arnab and Stojkoski, Viktor and Sandev, Trifce},
  booktitle={Target Search Problems},
  pages={323--355},
  year={2024},
  publisher={Springer}
}

@article{gupta2022stochastic,
  title={Stochastic Resetting: A (Very) Brief Review},
  author={Gupta, Shamik and Jayannavar, Arun M},
  journal={Frontiers in Physics},
  pages={130},
  year={2022},
  publisher={Frontiers}
}

@article{pal2022inspection,
  title={The inspection paradox in stochastic resetting},
  author={Pal, Arnab and Kostinski, Sarah and Reuveni, Shlomi},
  journal={Journal of Physics A: Mathematical and Theoretical},
  volume={55},
  number={2},
  pages={021001},
  year={2022},
  publisher={IOP Publishing}
}

@article{nagar2023stochastic,
  title={Stochastic resetting in interacting particle systems: A review},
  author={Nagar, Apoorva and Gupta, Shamik},
  journal={Journal of Physics A: Mathematical and Theoretical},
  year={2023},
  publisher={IOP Publishing}
}

@article{kusmierz2014first,
  title={First order transition for the optimal search time of L{\'e}vy flights with resetting},
  author={Kusmierz, Lukasz and Majumdar, Satya N and Sabhapandit, Sanjib and Schehr, Gr{\'e}gory},
  journal={Physical Review Letters},
  volume={113},
  number={22},
  pages={220602},
  year={2014},
  publisher={APS}
}

@article{pal2016diffusion,
  title={Diffusion under time-dependent resetting},
  author={Pal, Arnab and Kundu, Anupam and Evans, Martin R},
  journal={Journal of Physics A: Mathematical and Theoretical},
  volume={49},
  pages={225001},
  year={2016}
}

@article{pal_first_2019,
	title = {First passage under stochastic resetting in an interval},
	volume = {99},
	journal = {Physical Review E},
	author = {Pal, Arnab and Prasad, V. V.},
	year = {2019},
	pages = {032123},
}

@article{das2026freezing,
  title={Freezing transition and shock-like transients in ballistic random walks under threshold resetting},
  author={Das, Rahul and Majumdar, Satya N and Pal, Arnab},
  journal={arXiv preprint arXiv:2611.16733},
  year={2026}
}

@article{REDNER2023,
  author        = {Redner, S.},
  title         = {A first look at first-passage processes},
  journal       = {Physica A},
  volume        = {631},
  pages         = {128545},
  year          = {2023},
  doi           = {10.1016/j.physa.2023.128545}
}

@article{Krapivsky_2018,
  author        = {Krapivsky, P L and Redner, S},
  title         = {First-passage duality},
  journal       = {J. Stat. Mech.},
  volume        = {2018},
  number        = {9},
  pages         = {093208},
  year          = {2018},
  doi           = {10.1088/1742-5468/aaddb3}
}

@article{condamin2007first,
  author        = {Condamin, S. and B\'enichou, O. and Tejedor, V. and Voituriez, R. and Klafter, J.},
  title         = {First-passage times in complex scale-invariant media},
  journal       = {Nature},
  volume        = {450},
  number        = {7166},
  pages         = {77--80},
  year          = {2007},
  doi           = {10.1038/nature06201}
}

@book{schuss2013brownian,
  author        = {Schuss, Zeev},
  title         = {{Brownian} Dynamics at Boundaries and Interfaces: In Physics, Chemistry, and Biology},
  series        = {Applied Mathematical Sciences},
  volume        = {181},
  publisher     = {Springer},
  address       = {New York},
  year          = {2013},
  doi           = {10.1007/978-1-4614-7687-0},
  isbn          = {9781461476870}
}

@article{bray2013persistence,
  author        = {Bray, Alan J. and Majumdar, Satya N. and Schehr, Gr\'egory},
  title         = {Persistence and first-passage properties in nonequilibrium systems},
  journal       = {Adv. Phys.},
  volume        = {62},
  number        = {3},
  pages         = {225--361},
  year          = {2013},
  doi           = {10.1080/00018732.2013.803819}
}

@article{holcman2014narrow,
  author        = {Holcman, D. and Schuss, Z.},
  title         = {The Narrow Escape Problem},
  journal       = {SIAM Rev.},
  volume        = {56},
  number        = {2},
  pages         = {213--257},
  year          = {2014},
  doi           = {10.1137/120898395}
}

@article{redner1999capture,
  author        = {Redner, S. and Krapivsky, P. L.},
  title         = {Capture of the lamb: Diffusing predators seeking a diffusing prey},
  journal       = {Am. J. Phys.},
  volume        = {67},
  number        = {12},
  pages         = {1277--1283},
  year          = {1999},
  doi           = {10.1119/1.19115}
}

@article{evans2011diffusion,
  author        = {Evans, Martin R. and Majumdar, Satya N.},
  title         = {Diffusion with Stochastic Resetting},
  journal       = {Phys. Rev. Lett.},
  volume        = {106},
  number        = {16},
  pages         = {160601},
  year          = {2011},
  doi           = {10.1103/physrevlett.106.160601}
}

@article{evans2011prl,
  author        = {Evans, Martin R and Majumdar, Satya N},
  title         = {Diffusion with optimal resetting},
  journal       = {J. Phys. A: Math. Theor.},
  volume        = {44},
  number        = {43},
  pages         = {435001},
  year          = {2011},
  doi           = {10.1088/1751-8113/44/43/435001}
}

@article{evans2020stochastic,
  author        = {Evans, Martin R and Majumdar, Satya N and Schehr, Gr\'egory},
  title         = {Stochastic resetting and applications},
  journal       = {J. Phys. A: Math. Theor.},
  volume        = {53},
  number        = {19},
  pages         = {193001},
  year          = {2020},
  doi           = {10.1088/1751-8121/ab7cfe}
}

@article{pal2017first,
  author        = {Pal, Arnab and Reuveni, Shlomi},
  title         = {First Passage under Restart},
  journal       = {Phys. Rev. Lett.},
  volume        = {118},
  number        = {3},
  pages         = {030603},
  year          = {2017},
  doi           = {10.1103/physrevlett.118.030603}
}

@article{reuveni2016optimal,
  author        = {Reuveni, Shlomi},
  title         = {Optimal Stochastic Restart Renders Fluctuations in First Passage Times Universal},
  journal       = {Phys. Rev. Lett.},
  volume        = {116},
  number        = {17},
  pages         = {170601},
  year          = {2016},
  doi           = {10.1103/physrevlett.116.170601}
}

@article{pal2019landau,
  author        = {Pal, Arnab and Prasad, V. V.},
  title         = {{Landau}-like expansion for phase transitions in stochastic resetting},
  journal       = {Phys. Rev. Research},
  volume        = {1},
  number        = {3},
  pages         = {032001(R)},
  year          = {2019},
  doi           = {10.1103/physrevresearch.1.032001}
}

@article{Biswas2025_Threshold,
  author        = {Biswas, Arup and Majumdar, Satya N. and Pal, Arnab},
  title         = {Target Search Optimization by Threshold Resetting},
  journal       = {Phys. Rev. Lett.},
  volume        = {135},
  number        = {22},
  pages         = {227101},
  year          = {2025},
  doi           = {10.1103/752c-wqly}
}

@article{burkitt2006review,
  author        = {Burkitt, A. N.},
  title         = {A Review of the Integrate-and-fire Neuron Model: I. Homogeneous Synaptic Input},
  journal       = {Biol. Cybern.},
  volume        = {95},
  number        = {1},
  pages         = {1--19},
  year          = {2006},
  doi           = {10.1007/s00422-006-0068-6}
}

@article{debruyne2020optimization,
  author        = {De Bruyne, B. and Randon-Furling, J and Redner, S.},
  title         = {Optimization in First-Passage Resetting},
  journal       = {Phys. Rev. Lett.},
  volume        = {125},
  number        = {5},
  pages         = {050602},
  year          = {2020},
  doi           = {10.1103/physrevlett.125.050602}
}

@article{fuchs2016stochastic,
  author        = {Fuchs, Jaco and Goldt, Sebastian and Seifert, Udo},
  title         = {Stochastic thermodynamics of resetting},
  journal       = {Europhys. Lett.},
  volume        = {113},
  number        = {6},
  pages         = {60009},
  year          = {2016},
  doi           = {10.1209/0295-5075/113/60009}
}

@article{gupta2020workfluctuations,
  author        = {Gupta, Deepak and Plata, Carlos A. and Pal, Arnab},
  title         = {Work Fluctuations and {Jarzynski} Equality in Stochastic Resetting},
  journal       = {Phys. Rev. Lett.},
  volume        = {124},
  number        = {11},
  pages         = {110608},
  year          = {2020},
  doi           = {10.1103/physrevlett.124.110608}
}

@book{Olver1974,
  author        = {Olver, Frank W. J.},
  title         = {Asymptotics and Special Functions},
  publisher     = {Academic Press},
  address       = {New York},
  year          = {1974},
  doi           = {10.1016/C2013-0-11254-8},
  isbn          = {9780125258500}
}

@article{3bc4-nlkt,
  author        = {Najeeb, Fazil and Pal, Arnab and Prasad, V. V.},
  title         = {Diffusion in a wedge geometry: First-passage statistics under stochastic resetting},
  journal       = {Phys. Rev. E},
  volume        = {113},
  number        = {2},
  pages         = {024104},
  year          = {2026},
  doi           = {10.1103/3bc4-nlkt}
}

@article{dy2008,
  author        = {Dy, Diandrew Lexter L. and Esguerra, J. P.},
  title         = {First-passage-time distribution for diffusion through a planar wedge},
  journal       = {Phys. Rev. E},
  volume        = {78},
  number        = {6},
  pages         = {062101},
  year          = {2008},
  doi           = {10.1103/physreve.78.062101}
}

@article{krapivsky2022,
  author        = {Krapivsky, P. L.},
  title         = {Diffusion in a fluid flow generated by a source at the apex of a wedge},
  journal       = {Phys. Rev. Fluids},
  volume        = {7},
  number        = {6},
  pages         = {064502},
  year          = {2022},
  doi           = {10.1103/physrevfluids.7.064502}
}

@article{majumder2015,
  author        = {Chupeau, Marie and B\'enichou, Olivier and Majumdar, Satya N.},
  title         = {Survival probability of a {Brownian} motion in a planar wedge of arbitrary angle},
  journal       = {Phys. Rev. E},
  volume        = {91},
  number        = {3},
  pages         = {032106},
  year          = {2015},
  doi           = {10.1103/physreve.91.032106}
}

@article{biswas2026optimal,
  author        = {Biswas, Arup and Majumdar, Satya N and Pal, Arnab},
  title         = {Optimal threshold resetting in collective diffusive search},
  journal       = {J. Phys. A: Math. Theor.},
  volume        = {59},
  number        = {38},
  pages         = {385002},
  year          = {2026},
  doi           = {10.1088/1751-8121/aea1ae}
}

@misc{das2026non,
  author        = {Das, Rahul and Majumdar, Satya N and Pal, Arnab},
  title         = {Non-equilibrium dynamics of drift-diffusion process under threshold resetting},
  year          = {2026},
  eprint        = {2609.24438},
  archiveprefix = {arXiv},
  primaryclass  = {cond-mat.stat-mech}
}

@article{garcia2026target,
  author        = {Garc{\'\i}a-Valladares, Gregorio and Prados, Antonio and Manacorda, Alessandro and Plata, Carlos A.},
  title         = {How target distributions shape optimal stochastic resetting},
  journal       = {Phys. Rev. E},
  volume        = {113},
  number        = {2},
  pages         = {024107},
  year          = {2026},
  doi           = {10.1103/9m8d-yjhw}
}

@article{biroli2026first,
  author        = {Biroli, Marco and Majumdar, Satya N. and Schehr, Gr\'egory},
  title         = {First-passage resetting gas},
  journal       = {Europhys. Lett.},
  volume        = {153},
  number        = {3},
  pages         = {31002},
  year          = {2026},
  doi           = {10.1209/0295-5075/ae3f4f}
}

@article{kracikova2013threshold,
  author        = {Kracikova, M and Akiri, G and George, A and Sachidanandam, R and Aaronson, S A},
  title         = {A threshold mechanism mediates {p53} cell fate decision between growth arrest and apoptosis},
  journal       = {Cell Death Differ.},
  volume        = {20},
  number        = {4},
  pages         = {576--588},
  year          = {2013},
  doi           = {10.1038/cdd.2012.155}
}

@article{dy2,
  author        = {Dy, Diandrew Lexter and Esguerra, Jose Perico},
  title         = {First-passage characteristics of biased diffusion in a planar wedge},
  journal       = {Phys. Rev. E},
  volume        = {88},
  number        = {1},
  pages         = {012121},
  year          = {2013},
  doi           = {10.1103/PhysRevE.88.012121}
}

@article{ray2021resetting,
  author        = {Ray, Somrita and Reuveni, Shlomi},
  title         = {Resetting transition is governed by an interplay between thermal and potential energy},
  journal       = {J. Chem. Phys.},
  volume        = {154},
  number        = {17},
  pages         = {171103},
  year          = {2021},
  doi           = {10.1063/5.0049642}
}

@article{entropy_resetting,
  author        = {Mori, Francesco and Olsen, Kristian St{\o}levik and Krishnamurthy, Supriya},
  title         = {Entropy production of resetting processes},
  journal       = {Phys. Rev. Research},
  volume        = {5},
  number        = {2},
  pages         = {023103},
  year          = {2023},
  doi           = {10.1103/physrevresearch.5.023103}
}

@book{Feller1971,
  author        = {Feller, William},
  title         = {An Introduction to Probability Theory and Its Applications},
  edition       = {2},
  volume        = {2},
  publisher     = {John Wiley \& Sons},
  address       = {New York},
  year          = {1971},
  isbn          = {9780471257097}
}

@misc{chen20262d,
  author        = {Chen, Kevin and Bressloff, Paul C.},
  title         = {{2D} target search with boundary-induced resetting},
  year          = {2026},
  eprint        = {2609.31085},
  archiveprefix = {arXiv},
  primaryclass  = {cond-mat.stat-mech}
}

@book{Doetsch1974,
  author        = {Doetsch, Gustav},
  title         = {Introduction to the Theory and Application of the {Laplace} Transformation},
  publisher     = {Springer},
  address       = {Berlin},
  year          = {1974},
  doi           = {10.1007/978-3-642-65690-3},
  isbn          = {9783642656903}
}
\end{document}